\DeclareUnicodeCharacter{2009}{\,}
\documentclass[]{interact}

\usepackage{algorithm} 

\usepackage{algpseudocode} 
\usepackage{amsmath} 
\usepackage{graphicx} 
\usepackage{caption} 
\usepackage{subcaption}
\usepackage{booktabs}
\usepackage{multirow}
\usepackage{tabularx} 
\usepackage{array} 
\usepackage{multirow} 
\usepackage{epstopdf}
\DeclareUnicodeCharacter{2010}{-}

\usepackage[version=4]{mhchem}
\usepackage{natbib}
\bibpunct[, ]{(}{)}{;}{a}{}{,}
\renewcommand\bibfont{\fontsize{10}{12}\selectfont}
\usepackage{xfrac}

\theoremstyle{plain}

\theoremstyle{definition}

\theoremstyle{remark}

\begin{document}

\title{A Novel Algorithm for Digital Lithological Mapping: Case Studies in Sri Lanka's Mineral Exploration}

\author{
\name{R.~M.~L.~S.~Ramanayake\textsuperscript{a}, D.~C.~Dammage\textsuperscript{b}, I.~Z.~M.~Zumri\textsuperscript{b}, K.~A.~R.~S.~Rodrigo\textsuperscript{b}, A.~A.~P.~Perera\textsuperscript{b}, D.~Fernando\textsuperscript{b},
G.~M.~R.~I.~Godaliyadda\textsuperscript{b}, H.~M.~V.~R.~Herath\textsuperscript{b}, M.~P.~B.~Ekanayake\textsuperscript{b}, A.~Senaratne\textsuperscript{c}, Fadi Kizel\textsuperscript{d}}
\affil{\textsuperscript{a}Department of Electrical and Computer Engineering, Rutgers University, New Brunswick, NJ 08901, USA; \textsuperscript{b}Department of Electrical and Electronic Engineering, Faculty of Engineering, University of Peradeniya, Peradeniya 20400, Sri Lanka; \textsuperscript{c}Department of Geology, Faculty of Science, University of Peradeniya, Peradeniya 20400, Sri Lanka; \textsuperscript{d}Department of Mapping and Geoinformation Engineering, Civil and Environmental Engineering, Technion-Israel Institute of Technology, Haifa 3200003, Israel}
}

\maketitle

\begin{abstract}
Conventional manual lithological mapping (MLM) through field surveys are resource-extensive and time-consuming. Digital lithological mapping (DLM), harnessing remotely sensed spectral imaging techniques, provides an effective strategy to streamline target locations for MLM or an efficient alternative to MLM. DLM relies on laboratory-generated generic end-member signatures of minerals for spectral analysis. Thus, the accuracy of DLM may be limited due to the presence of site-specific impurities. A strategy, based on a hybrid machine-learning and signal-processing algorithm, is proposed in this paper to tackle this problem of site-specific impurities. In addition, a soil pixel alignment strategy is proposed here to visualize the relative purity of the target minerals. The proposed methodologies are validated via case studies for mapping of Limestone deposits in Jaffna, Ilmenite deposits in Pulmoddai and Mannar, and Montmorillonite deposits in Murunkan, Sri Lanka. The results of satellite-based spectral imaging analysis were corroborated with X-ray diffraction (XRD) and Magnetic Separation (MS) analysis of soil samples collected from those sites via field surveys. There exists a good correspondence between the relative availability of the minerals with the XRD and MS results. In particular, correlation coefficients of 0.8115 and 0.9853 were found for the sites in Pulmoddai and Jaffna respectively. 
\end{abstract}


\section{Introduction}

\label{section:introduction}
In recent years, spectral imaging (SI) has found applications in various domains, including climate and environmental monitoring \citep{henderson1997sar,giuliani2017live}, biodiversity studies \citep{Seto_Fleishman_Fay_Betrus_2004}, ecosystem analysis \citep{Yang_2005}, and food quality assessment \citep{bandara2020validation,weerasooriya2020transmittance}. Airborne SI offers significant advantages for remote sensing (RS) due to its non-intrusive nature and its capacity to simultaneously capture data across a broad area at multiple spectral wavelengths. These characteristics underpin its utility in diverse applications.

Prominent fields benefiting from RS include food and agriculture \citep{ekanayake2018semi}, ecology \citep{zhang2021advances}, hydrology \citep{chen2020comparative}, and mineralogy \citep{Kusuma_Ramakrishnan_Pandalai_2012}. Remote sensing proves invaluable in these domains, demonstrating its versatility and applicability. Within mineralogy, RS plays a significant role in a process known as 'mineral indication' \citep{yousefi2016mineral}, which involves the identification of minerals. In this context, remote sensing primarily focuses on the spectral and radiometric properties of minerals, as opposed to their chemical and physical characteristics. Decades of research in mineral indication have produced spectral libraries used for mineral identification and classification, facilitating the generation of mineral distribution maps. The development in remote sensing (RS) techniques, as evidenced by research \citep{black2016automated,Gersman_Ben‐Dor_Beyth_Avigad_Abraha_Kibreab_2008,grebby2011integrating}, has opened up opportunities to enhance lithological mapping processes, thereby improving our ability to interpret the compositions of geological sites. 

In this context, both hyperspectral images (HSIs) and multispectral images (MSIs) acquired through wide and narrow band sensors \citep{del2020evaluation} play vital roles in advancing digital lithological mapping (DLM). More specifically, within the field of Digital Lithological Mapping (DLM), Thermal Infrared (TIR) data obtained from strategically positioned multispectral sensors have proven to be exceptionally valuable. Notable examples include sensors like the Thermal Infrared Multispectral Scanner and the Advanced Space-borne Thermal Emission and Reflection Radiometer (ASTER), which have showcased the remarkable capability of TIR data in effectively discriminating a wide variety of minerals, with a particular emphasis on silicates.\citep{hubbard2005mineral,jiang2013lithological,Bishop_Liu_Mason_2011}.

Multispectral images (MSIs) are favored for DLM due to their inherent advantage of an improved signal-to-noise ratio (SNR) resulting from wide spectral band spacing. This spacing reduces cross-channel interference compared to hyperspectral images (HSIs). However, HSIs can be employed with algorithms that provide superior performance, even amidst noise interference, owing to their spectral richness. It's worth noting that the study sites in these investigations do not solely consist of minerals, devoid of other environmental features such as trees, soil, sand, etc. This factor makes it challenging to directly apply laboratory-generated spectral signatures \citep{ismail2014rare,kruse2007regional,pour2019lithological}.

Alternatively, spectral libraries \citep{Shanmugam_SrinivasaPerumal_2014} can be employed to identify a group of known minerals. This is achieved by applying criteria such as maximum likelihood classification \citep{cabral2018burned}, spectral angle mapping \citep{dennison2004comparison}, and spectral information divergence \citep{palsson2017neural}. These techniques prove effective because even if there is a minor error in identifying a mineral, it doesn't lead to significant issues since the error is distributed across various mineral types. However, these same methods may not perform well when the goal is to pinpoint a specific mineral. This is due to the fact that when a mineral is mixed with other substances, its spectral characteristics undergo changes, rendering it distinct from what's available in the library. 

In this study, a novel algorithm is introduced to generate the map of a target mineral of interest under the influence of various impurities. The algorithm consists of extracting a mineral representative signal from the HSIs, which will then be used to generate a map of the relative availability of the mineral in a given location. In contrast to the more traditional heuristic methods, one of the main steps in the algorithm is the autonomous separation of the mineral and the impurity signals from the HSI into the mineral and impurity representative subclasses. This subsequently opens up a way to utilise  Fisher's Discriminant Analysis, which maximises the separability of these two subclasses. As a result of this maximisation of separability, a quantity defined as the relative availability of a specific target mineral in a given HSI pixel is calculated. Finally, by utilising this quantity, a digital map of the mineral is generated.    

In order to firmly validate the performance of the algorithm, it was applied to three different minerals in four regions of Sri Lanka. The study considered Jaffna, Mannar, Pulmoddai, and Murunkan, all of which are locations in the Northern parts of Sri Lanka where mineral excavations are ongoing. The Jaffna area is popular for its Limestone, whereas Mannar and Pulmoddai both of which are well known for Ilmenite, and finally, Murunkan which is mostly famous for its clay like mineral, Montmorillonite. An extensive and rigorous study was conducted to compare the XRD and Magnetic Separation test results of the collected samples from these locations with the predicted availability of each site from the algorithm. The results demonstrate that the algorithm is capable of providing information with regards to the availability of a target mineral in a given location amidst impurities. 

Thus, the main contributions of this paper are,
\begin{itemize}
    \item A novel hybrid machine-learning and signal-processing based algorithm for identifying and mapping any target mineral, utilising remotely sensed hyperspectral imaging, regardless of the available mineral concentration.
    \item A case study that validates the results of the proposed algorithm, corroborating them with field surveys along with laboratory XRD and MS testing, is also presented. The case field survey data were collected spanning four sites (Mannar, Pulmoddai, Jaffna and Murunkan) and three minerals (Ilmenite, Limestone and Montmorillonite).
\end{itemize}
In addition, the following outcomes of the study are presented as supplementary contributions.
\begin{itemize}
    \item In order to maximize the separation between the mineral and the impurities, a vector in the feature space onto which the soil pixels can be projected has been calculated for each of the three minerals considered in the study.  
    
    \item The creation of a mineral map using the unique mineral signature of each site as a predictive model for determining which sites to include or skip during the field survey.

    \item Addressing the reduced accuracy resulting from solely relying on laboratory reference signatures for estimating mineral abundances by introducing a region-specific mineral representative signature.    
\end{itemize}

\section{Related work}
\label{section: related work}

The introduction of hyperspectral sensors with multiple spectral channels has significantly improved the accuracy of remotely retrieved mineralogical and surface composition information \citep{niranjan2016mapping,swamy2017remote,zomer2009building}. These sensors cover a wide spectral range, including the Thermal Infrared Region (TIR), and instruments like the Airborne Hyperspectral Scanner and others have been successful in mapping various minerals \citep{vaughan2003sebass,zhang2018hyperspectral}. Hyperion sensor data from the EO1 satellite has effectively mapped lithological units in India \citep{rs12010177}. Pseudo-hyperspectral imagery, generated from Landsat data and regression-based models, has been used to identify metal deposit-related minerals \citep{nguyen}. Hyperspectral images from satellites and airborne systems have also been employed to analyze minerals and rock microstructures while considering soil variability at different remote sensing scales \citep{van2019measuring,hecker2019spectral}.

In most digital lithological studies \citep{black2016automated,ekanayake2019mapping,Pal_Majumdar_Bhattacharya_Bhattacharyya_2011,grebby2011integrating,tziolas2020integrated},the prevalent approach has been the utilization of generic laboratory-generated spectral libraries, with limited emphasis on generating site-specific signatures. Furthermore, when employing generic spectral signatures for lithological mapping, assumptions were often made regarding the presence of minerals at the site \citep{ninomiya2019thermal,yu2012towards} or relied on pre-existing surveys of the site to confirm mineral composition \citep{pour2018mapping,xiong2011lithological}. The former approach can be error-prone due to the lack of site-specificity in the signatures, while the latter serves primarily as a validation procedure, offering neither enrichment nor acceleration of the field survey \citep{rajan2019mapping}.

The extraction of endmember signatures is often addressed and is a discrete research area in the literature for RS unmixing. These studies make use of unsupervised unmixing techniques in RS to derive the endmember signatures and the abundance map for the lithological map. For example, orthogonal subspace projection methods \citep{cheng2015remote,li2015recursive,ren2000generalized} were used to classify spectral signatures following dimension reduction algorithms. Similarly, several dimensionality reduction-based methods such as PCA, ICA, and MNF have been applied in mineral exploration with ASTER satellite data \citep{articlehojat}. Furthermore, accurate HSI classification has been performed via an improved version of the standard non-negative matrix factorization (NMF) algorithm incorporating fundamental notions of independence \citep{benachir2013hyperspectral,sun2017poisson}. 

Though existing techniques such as NMF-based unmixing \citep{rajabi2013hyperspectral,rathnayake2020graph,wang2016hypergraph} or autoencoder architecture-based unmixing \citep{hua2020autoencoder,Khajehrayeni_Ghassemian_2020,ranasinghe2020convolutional} are superior at extracting the endmembers and estimating the corresponding abundances, these algorithms require knowledge about the number of endmembers to extract which was not available in the first place. Besides, for an algorithm to find feasible survey locations for a particular mineral through single-target identification, such information is superfluous. Similarly, the performance of blind source separation algorithms, such as the Pixel-Purity-Index salgorithm and Independent Component Analysis (ICA), which are used to extract sources of signals will be hindered due to the lack of pure mineral pixels and information on the endmembers. However, most of the study areas considered had a mineral composition that allowed accurate endmember extraction via automated unmixing algorithms. While the endmember extraction methods mentioned above are suitable for unpopulated geographical locations, in certain geographical regions, human intervention \citep{barbosa2003otter,mendes2015multi,vila2016stakeholder} has altered the composition of the soil with impurities. This alteration could make these methods unreliable in these areas.

The availability of high-resolution remote sensing data has led to the accumulation of vast amounts of big data, opening new opportunities for data-driven discoveries. Currently, the integration of remotely sensed data with state-of-the-art data analytics, including machine learning, is significantly enhancing the field of geological mapping \citep{bachri}. For instance, certain remote sensing (RS) techniques and enhancement filters have been employed to trace geological structures and map hydrothermal alteration minerals using SPOT-5 and ASTER satellite data for Cu-Au prospecting \citep{Ahmadirouhani}. Additionally, semi-automated support vector machines were  utilised to map lithium (Li)-bearing pegmatites \citep{Ahmadirouhani}, and the random forest classifier was applied for lithological classification based on remote sensing data \citep{bachri}.

According to the related works, DLM has been performed using spectral libraries of geographical regions that had a pure mineral composition, thereby the mineral signatures were more dominant than the spectral signatures of other non-mineral constituents or impurities. Although there is a limited number of studies to improve the SNR of remote sensing images, the detection of minerals under the strong influence of impurities is yet to be improved to the best of our knowledge. On the other hand, estimation of the abundances for mineral mapping has been performed under multi-target detection with unmixing techniques. But the mineral map generation for a single mineral has not been under scrutiny since these unmixing techniques are susceptible to noise in the image. 

Addressing these gaps in the literature, this study introduces a comprehensive algorithm designed for the single-target detection of any mineral of interest in regions characterized by high impurity levels. The algorithm leverages Hyperspectral Images (HSIs) due to their superior spectral discrimination capabilities, despite their inherent challenges such as low SNR arising from cross-channel interference, as compared to Multispectral Images (MSIs) (\citep{guo2023stereo,zhao2017heterogeneous}. Furthermore, the proposed algorithm makes use of existing hyperspectral libraries for single-target mineral detection. Thus, this research addresses the identified deficiencies in the field, as outlined in Section \label{section:introduction}, by contributing site-specific endmember generation techniques aimed at estimating the abundances of a specific mineral in the context of single-target detection.

\section{Study Locations and Minerals}
\label{section: Study Locations and Minerals}

\subsection{Montmorillonite}
\label{subsection: Montmorillonite}

Montmorillonite which is of the chemical composition \ce{Al2H2O12Si4}  is a subclass of smectite and a natural aluminomagnusium silicate clay. The clay is of a three-layered crystalline structure \ce{(Na,Ca)_{0.3}(Al,Mg)_{2}Si_{4}O_{10}(OH)_{2}$\cdot$ nH_{2}O} in which silicate layers sandwich a layer of  aluminium oxide \citep{wanasinghe2017extraction}. The uses of Montmorillonite can be found in many different disciplines. It can significantly improve the needed performance when added to materials, polymers, and goods. It's interesting to note how widely Montmorillonite is used in polymers and composites as a functional filler. \citep{herath1973industrial} Furthermore, it is used as a food additive for health and stamina, antibacterial activity, and as a sorbent for nonionic, anionic, and cationic dyes catalysts in organic synthesis, etc. In addition, it contributes as a plasticizer in sand casting, in drilling mud, and in electrical, heat, and acid-resistant porcelain.

The prior research conducted on the distribution of different clay minerals in Sri Lanka provides evidence of Montmorillonite rich-clay deposits in the regions of Murunkan in Mannar, the vicinity around the Giant's tank \citep{wanasinghe2017extraction}. The Murunkan region is characterized by its cracked earth, giving it its name, which means "cracked earth" in Tamil. The name is derived from the black clay surface that tends to crack in hot, dry weather. 

The clay deposits are abundant in the dry region and are concentrated around the water reservoirs which have been built as sources of water for cultivation. In fact, the prior geological research state that Montmorillonite is accumulated at the bottom of the reservoirs due to sedimentation \citep{herath1973industrial}. This study focuses on the region of Giant’s Tank Sanctuary and Murugan in Mannar. The samples were collected from three sites; in the vicinity of the Giant tank, and the paddy fields on the roadsides of the Madawachchiya-Talaimannar highway.

\subsection{Limestone}
\label{subsection:Limestone}
Limestone, with the chemical composition of \ce{CaCO3}, is considered to be a malleable rock. Although Limestone is a sedimentary rock and is typically grey in colour, certain climatic and geographic factors can cause it to appear brown, white, or even yellow. It is extensively used as a major ingredient in cement production, road construction, and decorative designs. Furthermore, it is used in numerous medical applications and has scarce uses in the sculpture industry due to its porosity and softness.

The East coast of the Jaffna Peninsula of Sri Lanka is such a geographical site where the lithological data of the dominant mineral is unavailable. The thick Jaffna Limestone is the dominant rock type that underlies the whole of the Jaffna Peninsula \citep{senaratne1982palaeogeographic} and the surrounding islands. Formerly, Jaffna was an island composed of Miocene Limestone and the island was connected to the mainland by a spit formed of sediments brought by currents. These sediments were then carried to the eastern and northern coasts of the mainland and subsequently to the lagoon. The terrain of interest is a paleo-spit formed on the northeastern side of the peninsula which constitutes the shoreline, vegetation, and water bodies. However, the Limestone deposits are copious towards the West coast of Sri Lanka according to the findings of \citep{perera2020cement}. The study location selected in the work was in the vicinity of the Limestone region but lacks information about the existence of Limestone from previous studies.

\subsection{Ilmenite}
\label{subsection: Ilmenite}

Ilmenite or titanium-iron oxide (\ce{FeTiO3}) is a typical auxiliary mineral found in igneous and metamorphic rocks \citep{ilmenite1,ilmenite2}. It is mostly extracted from the ore sources of heavy mineral sands. The uses of minerals can be identified in many areas. In steel production, the Ilmenite ore is used as a flux to line the refractory in blast furnace hearths. Furthermore, Ilmenite can be used to produce ferrotitanium, which is necessary for the production of stainless steel, through an aluminothermic reduction. With the development of the industrial market, Ilmenite has emerged as one of the primary sources of titanium and titanium dioxide \citep{ilmenite1}, which are used to make paints, surface coatings, plastics, papers, and pharmaceutical substances.  

The proposed algorithm is used to identify the presence of Ilmenite in two geographical locations in Sri Lanka. Pulmoddai in the northeastern coastal region of the island is considered to be one of the richest mineral sand deposits\citep{ekanayake2019mapping}. This sand reserve is known for Ilmenite, Rutile, Leucoxene, Garnet, Monazite, Zircon, and Sillimanite, as well as minor amounts of Magnetite \citep{amalan2018influence}. The soil samples were collected from the sites near the coastline as well as in the vicinity of the distributors of Yan Oya. As the second region of interest, the southern coast of Mannar Island which is located on the northwest coastline of Sri Lanka was selected. Mannar has been identified as a potential area for heavy mineral placer deposits. It was observed that the geographical regions constituted mainly of vegetation, soil, paddy fields, beaches with the sand, sea, and other water bodies.

\begin{table}[h]
    \caption{Locations and Minerals}
    {\begin{tabular}{lccc}
        \toprule
        Image ID & Mineral of Interest & Locations of Interest \\ 
        \midrule
        \multirow{2}{*}{EO1H1410532005132110PX\_PF1\_01} & Limestone & Jaffna \\
                                                        & Montmorillonite & Murunkan (Mannar) \\ \\
        EO1H1420542007046110PY\_PF1\_01                 & Ilmenite & Mannar Island \\ \\
        EO1H1410532005260110PU\_SGS\_01                 & Ilmenite & Pulmoddai \\
        \bottomrule
    \end{tabular}}
    \label{tab:radiometric table}
\end{table}
\label{subsection: locations and minerals}
\subsection{Remotely Sensed Data}
\label{subsection: remotely sensed data}
\subsubsection{Hyperion Sensor and Its Details}
\label{Hyperion Sensor and Its Details}

\begin{figure*}[ht!]
    \centering \includegraphics[width=1.0\textwidth]{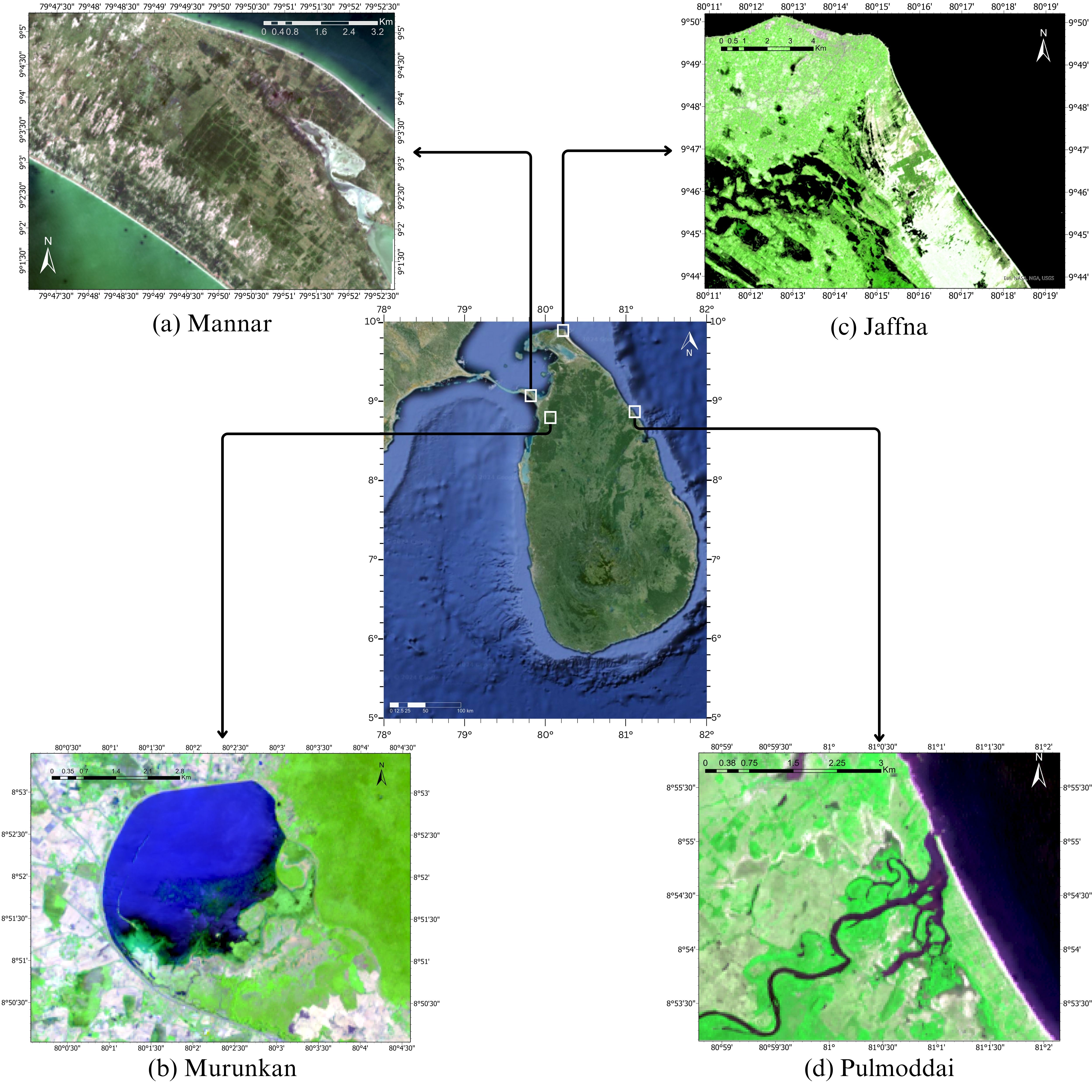}
        \caption{Study Locations}
        \label{figure: sri lanka}
\end{figure*}

The HSIs required for the study were captured through the Hyperion Sensor of “NASA’s Earth Observing 1” satellite. The spectral resolution of the instrument is 242 bands ranging from 0.4 to 2.5 um with a spatial resolution of 30 meters. For the data acquisition, the United States Geological Survey (USGS) database was used. The acquired images are represented in Fig. \ref{figure: sri lanka}. Conducting the analysis on the original data strip requires high computational power. Therefore, the regions of interest were cropped from the original HSI to minimise the computational burden.        

All the images captured from the Hyperion sensor are provided as Level 1Gst products which are terrain-corrected and available in the form of 16-bit radiance values. In order to perform further analysis on the product the radiance values were converted into reflectance values according to the equation proposed by \citep{kokaly2017usgs}. Ancillary data required for this purpose such as mean solar exo-atmospheric irradiance for each band, and earth-sun distance in astronomical units for a set of days of the year are available in the USGS database. The geometrical parameters used for the conversion are given in Table \ref{table: radiometric parameters}.

\begin{table}[h]
    \centering
    \caption{Parameters required for the radiometric conversion}
    \begin{tabular}{lccc}
        \toprule
        Image ID & Earth-sun distance & Solar Zenith Angle \\
        \midrule
        EO1H1410532005132110PX\_PF1\_01 & 1.010189776177688 & 28.543857 \\ \\
        EO1H1420542007046110PY\_PF1\_01 & 0.9876243413596166 & 37.385832 \\ \\
        EO1H1410532005260110PU\_SGS\_01 & 1.0051524576883322 & 27.520057 \\
        \bottomrule
    \end{tabular}
    \label{table: radiometric parameters}
    \vspace{-\baselineskip} 
\end{table}

\subsubsection{Extraction of Laboratory Reference Signature for each Mineral}
\label{subsubsection: Extraction of Laboratory Reference Signature for each Mineral}

The proposed algorithm requires a laboratory spectral signature of the explored mineral to use as the reference. These spectral signatures for the minerals; Ilmenite, Montmorillonite, and Limestone were obtained from the USGS Spectral Library \citep{kokaly2017usgs}. However, the laboratory spectral signatures of the minerals have been obtained through different sensing equipment. Therefore, an interpolation approach was carried out to promote continuity among the spectral bands. The details of the three instruments are tabulated in Table \ref{tab:wavelength}.

\begin{table}[h]
    \vspace{20pt}
    \centering
    \caption{Sensor Details}
    \label{tab:wavelength}
    \begin{tabular}{lccc}
        \toprule 
        Sensor                               & Wavelength ($\mu$m) & Number of Channels & Mineral                                                             \\ \midrule
        Hyperion                             & 0.4 - 2.5        & 242                & -                                                                   \\
         \multirow{2}{*}{Beckerman b}                          &  \multirow{2}{*}{0.2 - 3.0}        &  \multirow{2}{*}{480}                & Montmorillonite, \\
                                            &                   &                    & Ilmenite \\
        Analytic Spectral                &  \multirow{2}{*}{0.35 - 2.5}       &  \multirow{2}{*}{2151}               &  \multirow{2}{*}{Limestone} \\ 
        Devices Full Range                   &                  &                             & \\
        
        \bottomrule
    \end{tabular}
\end{table}

\section{Methodology}
\label{section: methodology}

\subsection{Pre Classification }
\label{subsection: pre-classification}

In many applications of hyperspectral imaging, it is vital to identify the pixels that correspond to a specific material or substance of interest. Specifically, this is crucial in applications such as lithological mapping. In many cases, as the mineral of interest is present in soil or sand as a minor component the algorithm proposes the isolation of the soil or sand representative pixels prior to doing further analysis on those pixels searching for the pixels with high mineral abundance. This hierarchical approach helps to prevent the misclassification of other materials as the mineral of interest while increasing the computational efficiency and accuracy of the algorithm. 

Remotely sensed HSIs contain both pure pixels that correspond to specific types of materials (such as water, vegetation, soil, sand, etc.) as well as mixed pixels. The specific endmembers of an HSI are influenced by the topography of the selected environment \citep{ref4} while the choice of the number of endmembers depends on the application and the complexity of hyperspectral data.  Most of the algorithms used in the literature for endmember extraction, require prior knowledge of the number of endmembers in the hyperspectral data \citep{ref1}. The methods of identifying the number of endmembers present in a scene include visual inspection \citep{ekanayake2019mapping} and statistical analysis \citep{ref3}.

To identify the number of dominant endmembers present in the HSI, initially, the Elbow method is used by iteratively changing the number of clusters through the K-means algorithm. The Elbow method is frequently used to find the optimal number of clusters when the class labels are unknown. For each HSI, the pixels were iteratively clustered while changing the number of clusters in each iteration (K in K-means clustering). At each iteration, the Within Cluster Sum of Squares (WCSS) is calculated according to Equation \ref{equation: similarity equation}. Then, the plots against WCSS and K are constructed. The optimum value for the K is selected by investigating the point where the graph experiences a significant change in the slope before it reaches the plateau.

Once the number of dominant endmembers has been identified, the HSI is pre-classified into sub-components. For this, Vertex Component Analysis (VCA) is used. First, the candidate endmembers were extracted using VCA for the number of classes suggested by the Elbow method. Then, each HSI pixel was normalized in the spectral direction using the L2-norm. Similarly, the candidate endmember signatures were also normalised. Thereafter, the Euclidean distance between the HSI spectra and each candidate end member was calculated. The final goal of this is to isolate the pixels corresponding to each candidate end member. Therefore, the reciprocal of the computed distance metric was used as a measure of affinity. The formula used for the similarity measure is computed through the following equation.

\begin{align}
    \label{equation: similarity equation}
    \gamma_\text{m}^\text{i} &= \frac{\sfrac{1}{\Vert\mathbf{u_m}-\mathbf{r_i}\Vert_2}}{\sum_i \sfrac{1}{\Vert\mathbf{u_m}-\mathbf{r_i}\Vert_2}}
\end{align}

where, $\mathbf{r_i}\text{ and } \mathbf{u_m}$ denote the spectral signatures of the i\textsuperscript{th} reference vector and the m\textsuperscript{th} pixel respectively, hence the notation of $\gamma_\text{m}^\text{i}$ for the similarity between m\textsuperscript{th} and i\textsuperscript{th} reference signatures. The summation is taken over all the candidate endmembers. If a pixel’s similarity with a candidate endmember exceeded 0.5, it was assigned to the respective candidate’s cluster. The pixels whose resultant affinity values were less than 0.5 were unassigned and disregarded. Once the pixels were classified, mineral identification was performed on the pixels classified as soil.

\subsection{Subclass Identification and Representative Endmember Extraction}
\label{subsection: mineral-residual impurity classification}
\begin{figure*}[ht!]
    \centering
    \begin{subfigure}[t]{0.9\textwidth}
        \centering
    \includegraphics[width=1.0\textwidth]{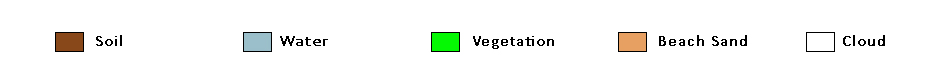}
    \end{subfigure}
    \medskip 
    \begin{subfigure}[t]{0.9\textwidth}
        \centering
\subcaptionbox{Jaffna}{\includegraphics[width=0.23\textwidth]{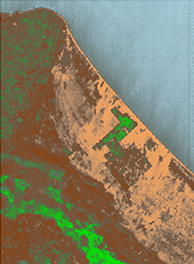}}%
\hfill 
\subcaptionbox{Pulmoddai}{\includegraphics[width=0.23\textwidth]{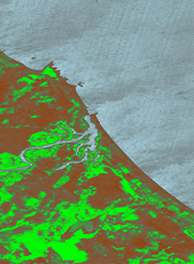}}%
\hfill 
\subcaptionbox{Mannar}{\includegraphics[width=0.23\textwidth]{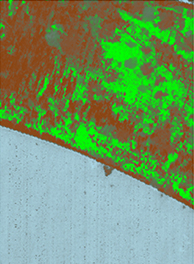}}%
\hfill 
\subcaptionbox{Giant's Tank}{\includegraphics[width=0.23\textwidth]{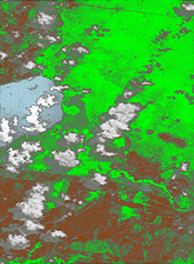}}%
    \end{subfigure}
    \caption{Generated Abundance Maps after pre-classification}
    \label{figure: abundances macro}
\end{figure*}

Once the pixels of the HSI are classified into the respective endmember classes based on their spectral properties, it can be seen that the classification perfectly highlights the land cover mapping of the given terrain (Fig. \ref{figure: abundances macro}). For the mineral representative pixel identification, the algorithm prefers the extraction of a mineral representative signature endemic to the scene from the HSI itself. As the targeted minerals are mostly present in soil, initially, the soil pixels were isolated from the others. The isolation of soil pixels before mineral-based classification reduces errors that arise due to misclassification, and confusion due to mixed pixels and reduces the computational cost. Next, the algorithm tries to identify the highly correlated pixels to the actual spectral signature of the target mineral. For this purpose, the laboratory mineral signatures for the three minerals Limestone, Montmorillonite, and Ilmenite were acquired from the USGS spectral library \citep{kokaly2017usgs}. The sensor details of the spectral library signatures and the mandatory preprocessing steps taken are mentioned in \ref{subsubsection: Extraction of Laboratory Reference Signature for each Mineral}.

Next, Correlation Factor Analysis was carried out using the HSI and the laboratory signature of the desired mineral. For this, the following equation which computes the Pearson’s Correlation Coefficient between two signals was used.

\begin{equation}
    \label{equation: pearson correlation calculation}
    r_k = \frac{\sum_{i=1}^b(x_{k,i}-\frac{1}{b}\sum_{j=1}^b x_{k,j})(s_i-\frac{1}{b}\sum_{j=1}^b s_j)}{\sqrt{{\sum_{i=1}^b}(x_{k,i}-\frac{1}{b}\sum_{j=1}^b x_{k,j})^2(s_i-\frac{1}{b}\sum_{j=1}^b s_j)^2}}
\end{equation}
where, \(r_k\), \(x_{k,n}\), \(s_n\), and \(b\) represent the Pearson correlation coefficient between \(k^{th}\) pixel and the reference signature, the value of the \(n^{th}\) spectral band of the \(k^{th}\) pixel, the value of the \(n^{th}\) spectral band of the laboratory Limestone signature, and the number of spectral bands, respectively.

In order to identify the highly correlated and the slightly correlated pixel classes with the mineral, threshold values for classification should be computed. Even though the computed correlation coefficients show the relationship between the soil signature with the ideal mineral signature, it might not reflect the relationship between the signature endemic to the specific site. However, this can be taken as an initial estimation for identifying pixels that tend to show closeness to the mineral at the site. Based on this, it was decided to create two classes, namely, the mineral representative class and the impurity representative class based on the computed correlation coefficients. In order to do this, it is necessary to identify the threshold values used to classify the pixels into the aforementioned classes. To compute these threshold values, initially, two endmembers were extracted from the soil pixel manifold, and the correlation of these pixels to the laboratory reference was computed. This approach was based on the hypothesis that, depending on the availability of the mineral in the area covered by a pixel, the pixel’s signature would exhibit inherent features. The authors try to extract two endmembers from the soil pixels using VCA. As VCA tries to separate two source signatures from the soil pixels, it was assumed that the mineral’s presence could be a differentiating factor when it comes to the extraction of the end members.

The signature that showed a high correlation to the reference mineral signature was chosen to be the mineral representative signature while the signature with a low correlation was assigned to represent the impurities naming it as the impurity representative signature. The threshold values for classification were taken to be the correlation coefficients of the above representative pixels.
\begin{itemize}
    \item Correlation Coefficient of Mineral Representative Signature ($C_1$)
    \item Correlation Coefficient of Impurity Representative Signature ($C_2$)
\end{itemize}

The pixels that showed a higher correlation than $C_1$ were assigned to the mineral representative sub-class, while the pixels that showed a lower correlation than $C_2$ were assigned to the impurity representative sub-class. Finally, the class representative signatures which were extracted from VCA prior to assigning the pixels into subclasses were replaced by the mean signature of each sub-class. The steps taken are summarized through the pseudo-code.

\begin{algorithm}[h!]
\caption{Separation of the Representative Classes}
\begin{algorithmic}[1]
\For{$\text{pixel} = 1$ \textbf{to} size of $HSI$}
    \If{$\text{pixel} == \text{soil pixel}$}
        \State $\text{corr}(pixel) = \text{Correlation}(HSI(pixel), mineral\_signature)$
    \Else
        \State Set $\text{corr}(pixel)$ to NaN
    \EndIf
\EndFor
\State Create ${soil\_matrix}$ by selecting soil pixels from the HSI 
\State Extract 2 endmembers by performing VCA on ${soil\_mat}$ and store as columns of ${rep\_sig}$ matrix.
\State Normalize extracted endmembers:
\State ${rep\_sig\_1} = \text{Normalize}(rep\_sig(:,1))$
\State ${rep\_sig\_2} = \text{Normalize}(rep\_sig(:,2))$
\State Calculate Pearson correlations between normalized endmembers and ${mineral\_signature}$:
\State $corr\_rep\_sig\_1 = \text{Correlation}(rep\_sig\_1, mineral\_signature)$
\State $corr\_rep\_sig\_2 = \text{Correlation}(rep\_sig\_2, mineral\_signature)$
\State Calculate lower and upper thresholds:
\State $lower = \text{Min}(corr\_rep\_sig\_1, corr\_rep\_sig\_2)$
\State $upper = \text{Max}(corr\_rep\_sig\_1, corr\_rep\_sig\_2)$ 
\For{$pixel = 1$ \textbf{to} size of $HSI$}
    \If{$corr(pixel) > upper$}
        \State Label $HSI(pixel)$ as $mineral\_representative$
    \ElsIf{$corr(pixel) < lower$}
        \State Label $HSI(pixel)$ as $impurity\_representative$
    \Else
        \State Label $HSI(pixel)$ as $not\_a\_soil\_pixel$
    \EndIf
\EndFor
\end{algorithmic}
\end{algorithm}

\subsection{Alignment of soil pixels for purity}
\label{subsection: alignment of soil pixels for purity}

To identify the mineral availability of the particular terrain and map the relative abundances of the earmarked mineral, it is necessary to investigate the correlation between the soil pixel and the mineral reference signature. This will allow one to identify patterns in the correlation variation when traversing from the least correlated set to the most correlated one along the soil signature manifold. To quantify the comparative availability of the mineral in the particular region of interest, we define a parameter named Relative Availability (RA)(Equation \ref{equation: relative availability}). This parameter is a normalised metric to measure the amount of minerals in a given soil pixel. In order to compute Relative Availability, the authors utilise a dimensionality reduction method. In the context of dimensionality reduction, several algorithms such as Principal component analysis, Singular Value Decomposition, t-SNE, and Fisher’s Discriminant Analysis (FDA) have been widely used\citep{dimered1,dimred2,dimred3}.  

However, to analyse the changing patterns of correlation when moving along the soil signature manifold conditioned on the mineral availability, it is possible to project the dataset in a direction that enhances the separation between the aforementioned classes. FDA is a statistical technique used to reduce dimensionality while promoting class discrimination. It tries to identify an eigen-direction that maximises the separation between classes. To be concise, it aims to reduce the projected within-class variation and simultaneously increase the gap between projected means. Within this framework, FDA was used to identify an eigen-direction with a clear separation between the two sub-classes.

Initially, FDA was performed on the two representative classes acquired in the previous step. Through this, the eigen-direction with the highest separation between the two classes was identified. Then, the pixel vectors belonging to the soil subclass were projected onto the new direction. Further, representative signatures of the two classes were also transformed onto the reduced one-dimensional space following the calculation of RA through the equation given below. Then, the distances to both representative signatures from the pixel were calculated for each pixel along the eigen-direction. The reciprocal of each distance was taken as a measurement \citep{fadi_projected_gradient} for the affinity, and the relative availability of the mineral was defined as,

\begin{equation}
    \label{equation: relative availability}
    \begin{split}
    \mathbf{\textit{Relative availability}} &= \frac{\textit{similarity with the mineral reference}}{\textit{total of similarities with references}}\\
    &= \frac{\sfrac{1}{d_m}}{\sfrac{1}{d_m}+\sfrac{1}{d_{i}}}\\
    &= \frac{d_{i}}{d_{m}+d_{i}}
    \end{split}
\end{equation}
where $d_{m}$, and $d_{i}$ are the distances from the pixel to the representative signatures of the mineral and impurity sub-classes, respectively.

\subsection{Least Square Estimation for Abundance Generation}
\label{least squares}
After calculating relative abundance values for the mineral of interest, the authors use least squares estimation to further enhance the calculated mineral abundance, assuming a linear mixture model between the mineral representative signature and the impurity representative signature. Initially, the mineral representative signature was replaced by the average signature of the pixels, emphasizing a relative availability greater than 0.8. Similarly, the impurity signature was replaced by the average signature of the pixels with a relative availability lower than 0.2. Assuming a linear mixture model the soil pixel signature, $\mathbf{s}$ can be represented as,

\begin{equation}
    \begin{split}
    \label{equation: formation of signature}
        \mathbf{s} &= \{\alpha\mathbf{m} + \beta\mathbf{r}~|~ 0\leq\alpha,\,\beta \leq 1;~ \alpha + \beta = 1\}\\
    \end{split}
\end{equation}

where $\mathbf{m}$ and $\mathbf{\alpha}$ denote the mineral representative signature and its coefficient, while $\mathbf{r}$ and $\mathbf{\beta}$ represent the impurity representative signature and its coefficient. The coefficients represent the fractions of each component in the soil pixel signature. By utilizing this formulation, the non-negative least squares estimation was used to calculate the optimal coefficients, $\mathbf{\alpha^*}$ and $\mathbf{\beta^*}$, of the linear mixture model.

In order to formulate the estimation of coefficients as a non-negative least squares problem, each pixel $\mathbf{s}$ is considered separately and estimated using the $\mathbf{A}$ matrix, which contains the two representative signatures (as column vectors). Furthermore, the vector $\mathbf{x}$ contains the two coefficients, $\mathbf{\alpha}$ and $\mathbf{\beta}$ as its elements. Therefore, the optimal vector $\mathbf{x^*}$ is found by minimizing the objective function with the constraints given below.
\\
\[
\begin{aligned}
& \text{Minimize} \quad \frac{1}{2} \left\| Ax - s \right\|_2^2
\end{aligned}
\]

Subject to the constraints:
\[
\begin{aligned}
& x \geq 0 \quad \text{(Element-wise non-negativity constraint)} \\
& \alpha + \beta = 1 \quad \text{(Sum-to-one constraint)}
\end{aligned}
\]
where, $\left\| A \mathbf{x} - \mathbf{s} \right\|_2^2$ represents the squared Euclidean norm of the residual error between the estimation ($\mathbf{Ax}$) and the pixel signature ($\mathbf{s}$), and the "2" as a subscript indicates the Euclidean norm. \(\mathbf{x}\) represents the vector which contains the coefficients $\alpha$ and $\beta$.

\subsection{Field Survey and Sample Collection}
\label{subsection: validation through site survey}

As outlined in section \ref{section: Study Locations and Minerals}, samples of Ilmenite, Limestone, and Montmorillonite were gathered from four distinct regions within Sri Lanka. After the algorithm had generated likely mineral availability locations, the authors, in coordination with the Department of Geology at the University of Peradeniya, Sri Lanka, conducted the site selection process. This selection was meticulously carried out, considering regions with both high and low mineral availability, aiming to validate the algorithm's performance.

\begin{figure}[H]
\centering
\subcaptionbox{}{\includegraphics[width=0.3\textwidth]{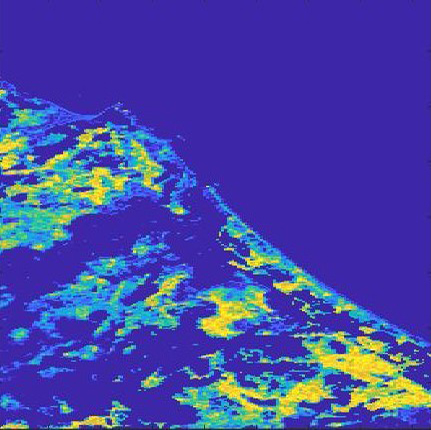}}%
\hfill 
\subcaptionbox{}{\includegraphics[width=0.3\textwidth]{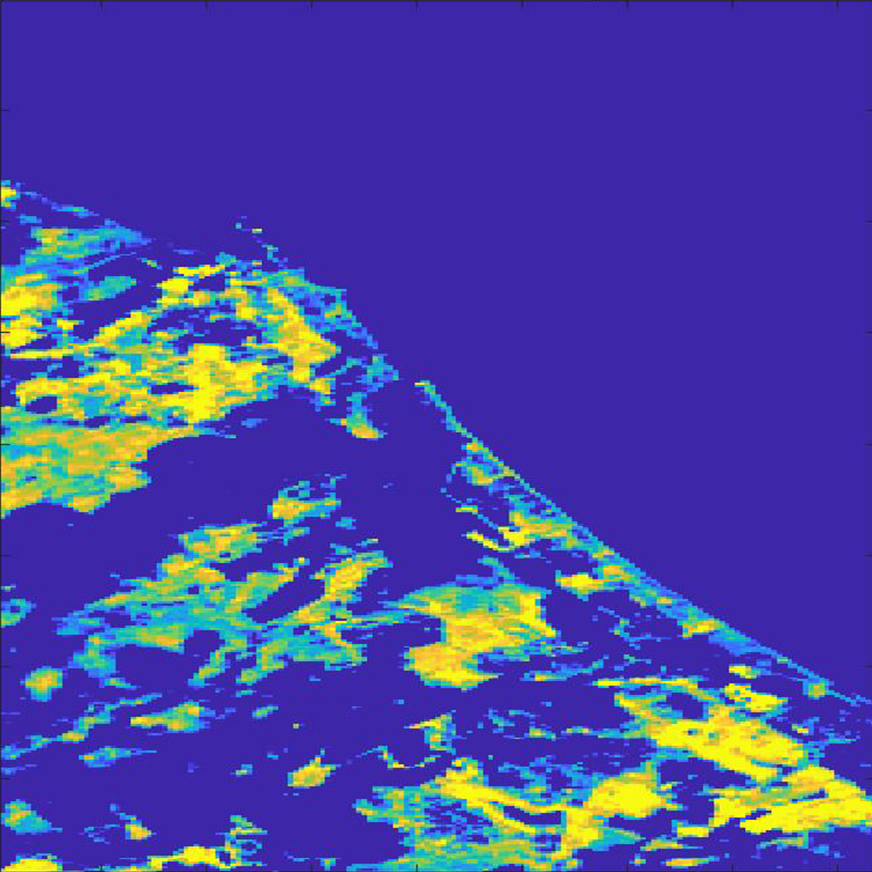}}%
\hfill 
\subcaptionbox{}{\includegraphics[width=0.3\textwidth]{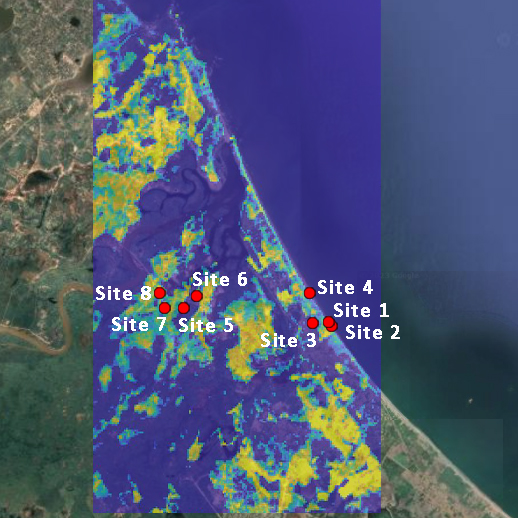}}%
\caption{(a) Relative availability Map (b) Abundance map generated after NNLS (Alpha values generated after NNLS) (c) Generated mineral map overlapped with sites for Pulmoddai}
\label{figure: pulmoddai relative availability}
\end{figure}

The mineral map generated in this study served as a valuable tool for identifying potential areas with high and low probabilities of containing the desired minerals. Once the pixels for the survey were selected from the generated mineral map, the precise GPS coordinates for the corresponding pixels were determined using ArcGIS. As it is impractical to collect samples from such a large area, the sample collection was carried out with the assumption that the soil is of a homogeneous composition. The representations of the selected sites for the sample collection are depicted in Figures \ref{figure: mannar relative availability} (c), \ref{figure: pulmoddai relative availability} (c), \ref{figure: jaffna relative availability} (c), \ref{figure:Giants-tank-relative-availability} (c).

    
Upon arriving at the sites of interest, soil samples were meticulously collected under the supervision of experienced members from the Department of Geology at the University of Peradeniya, Sri Lanka. Samples were obtained from various locations scattered across the sites of interest. During the collection process, extra care was taken to gently scrape the surface to gather samples without digging into the soil. This precaution is essential because remote sensing (RS) images solely capture material or information available on the surface.

\subsection{ Magnetic Separation and XRD for Mineral Identification}

The samples obtained were put under laboratory testing to examine the presence of the earmarked minerals. X-ray diffraction (XRD) test and Magnetic separation were carried out to identify the minerals. As the literature shows, the preferred testing procedure for Limestone and Montmorillonite \citep{wanasinghe2017extraction} is XRD, while for Ilmenite, it is magnetic separation. However, due to the texture and grain size of the samples collected from Mannar, it was decided to use the XRD test instead of the magnetic separation only to find the availability of Ilmenite in these soil samples. The mineral identification for the samples collected from Pulmoddai was done through magnetic separation using the FRANTZ Magnetic Separator Model L-1. Before testing for the mineral availability, each sample was prepared to adhere to the standard procedure with the aid of the Department of Geology, Faculty of Science, University of Peradeniya. The procedure followed is described in the following sections.

\subsubsection{Magnetic Seperation Method}
\label{magnetic separation}

Ilmenite is a mineral that possesses some magnetic susceptibility. Hence, it is possible to identify the availability of Ilmenite in collected specimens incorporating magnetic methods. This study used FRANTZ Magnetic Separator Model L-1 to separate Ilmenite. Prior to magnetic separation, the samples were prepared according to the steps described below.

\begin{enumerate}
    \item The soil samples were thoroughly washed with distilled water to remove impurities.
    \item After drying the samples for 24 hours under 100°C, they were passed through a set of sieves to separate them into different grain sizes.
    \item The standard particle size for magnetic separation is 75µm - 125µm. The samples with the preferred grain size were separated through the sieving process.
    \item A representative sample was selected from the separated sand, and the weight of the sample was measured.
    \item As the magnetic separator requires magnetite removal before the samples are entered into the separator, a hand-magnet was used to separate the magnetite mineral from the obtained soil samples.
    \item The remaining soil samples were inserted into the magnetic separator Model L-1 while the current was controlled at 0.4 A.
\end{enumerate}

Finally, Ilmenite particles were isolated from the separator. The Ilmenite availability was quantified as a percentage of the weight measured prior to magnetite removal.


\subsubsection{X-Ray Diffraction Method}
\label{XRD}

X-rays are generated due to the deceleration of electrically charged particles, which contain a sufficient amount of energy. The XRD test, which incorporates these, is considered to be a non-destructive technique that unveils the crystallographic structure, physical properties and chemical composition of a mineral. The availability of a particular mineral is detected through the analysis of diffracted rays, which are generated by directing collimated X-rays towards the sample. Before carrying out the XRD test, a pre-processing procedure unique to each mineral was followed. The main common steps followed for sample preparation are listed below.

\begin{enumerate}
    \item The soil samples were thoroughly washed with distilled water to remove impurities and dried for 24 hours at 100$^o$C in the oven.
    \item Representative samples were separated from each main sample from the sites.
    \item The representative soil samples were powdered using the mortar and pestle or the electrical agate.
    \item The powdered sample was filtered for 63 microns through the sieve shake.
    \end{enumerate}

Even though the commonly used testing procedure for Ilmenite is magnetic separation, in the raw soil sample, the amount of soil within the recommended grain size, 75$\mu$m - 125$\mu$m,  was naturally low. This made the representative sample size insufficient,  leading to extremely low Ilmenite separation from the magnetic separator. Because the validation for the proposed algorithm requires accurate results for the mineral availability, the authors decided to follow an XRD test, which is capable of tracing even a tiny amount of mineral available in a specimen.

For Limestone, the following analysis steps were conducted before sample preparation.

\begin{enumerate}
    \item The samples' texture, colour and grain size were manually inspected and logged.
    \item Reaction with hydrochloric acid was observed as \ce{CaCO3} in Limestone reacts with the acid.
    \item Samples were grouped as typical if the reaction with the acid is as expected, otherwise aberrant.
    \item Observations of the acid test were validated by checking for different compounds using a digital microscope.
    \item A total of twenty samples were selected (five samples from each site) for XRD. The selection was done by accounting for locations with high and low abundance values.
    \end{enumerate}

\begin{figure*}[ht!]
    \centering \includegraphics[width=1.0\textwidth]{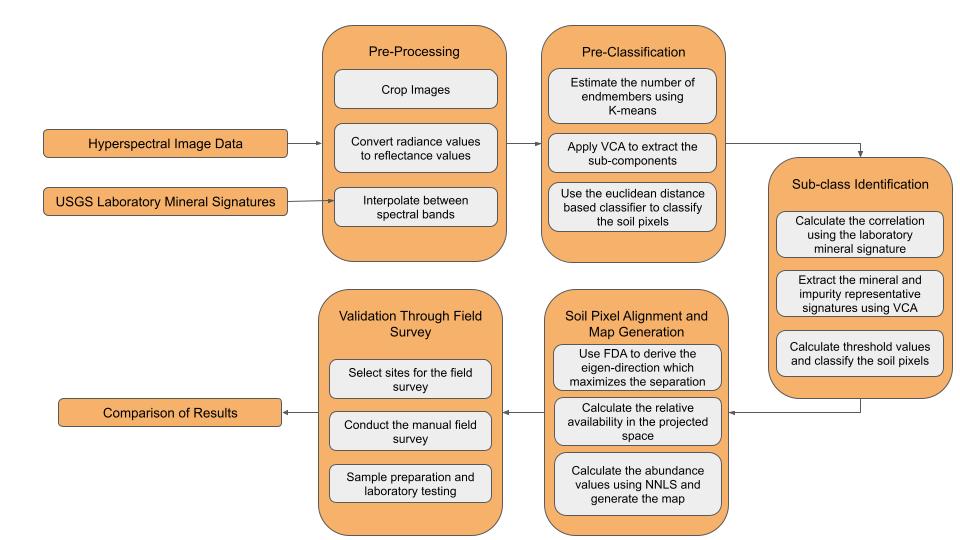}
        \caption{The Flow Chart Depicting the Methodology}
        \label{figure: flowchart}
\end{figure*}

\section{Results and Discussion}
\label{section: results}

\subsection{Identification of the Number of Endmembers}
\label{subsection: Identification of the Number of Endmembers}

\begin{figure}[H]
\centering
\subcaptionbox{Jaffna}{\includegraphics[width=0.25\textwidth]{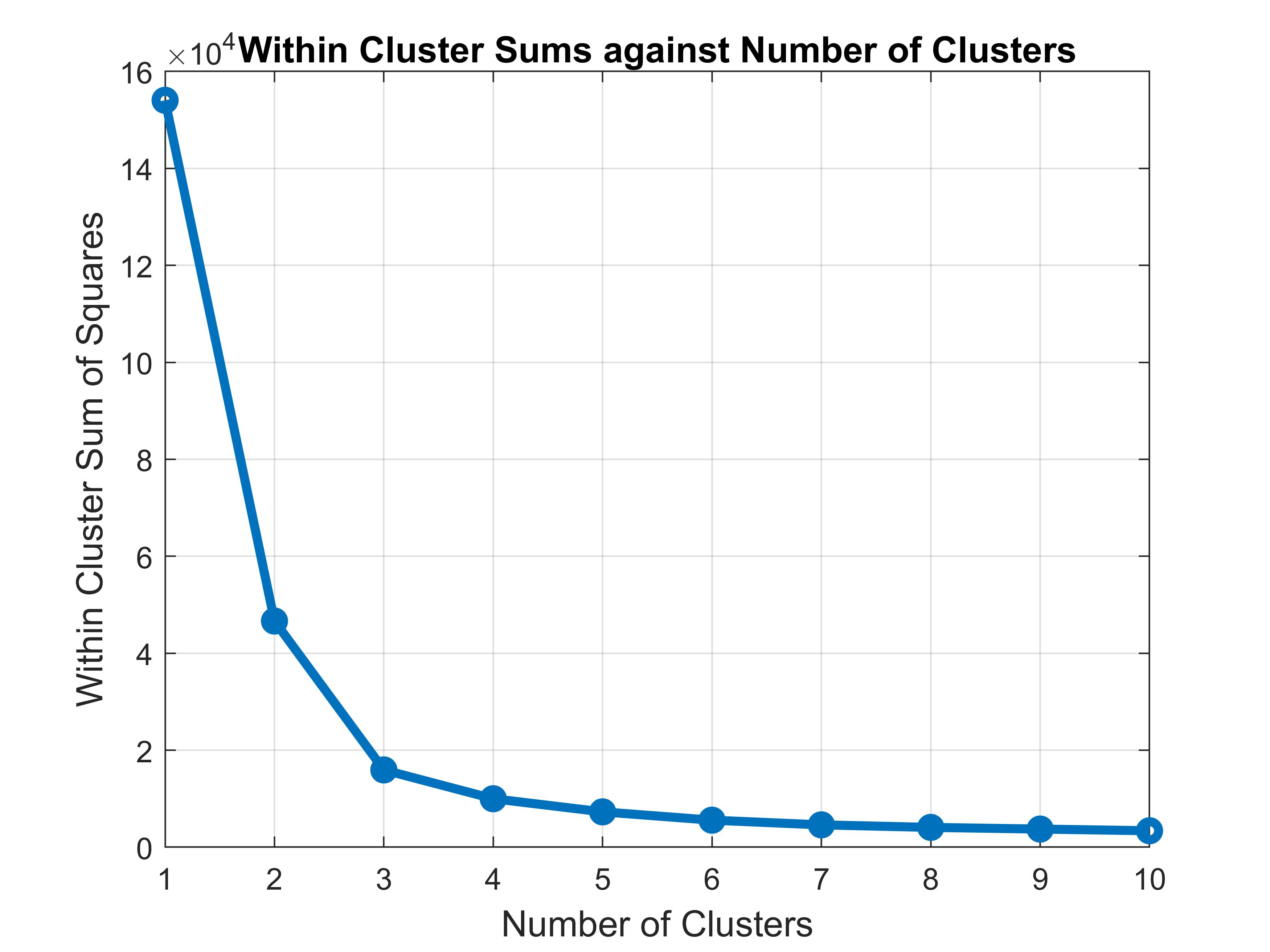}}%
\hfill 
\subcaptionbox{Pulmoddai}{\includegraphics[width=0.25\textwidth]{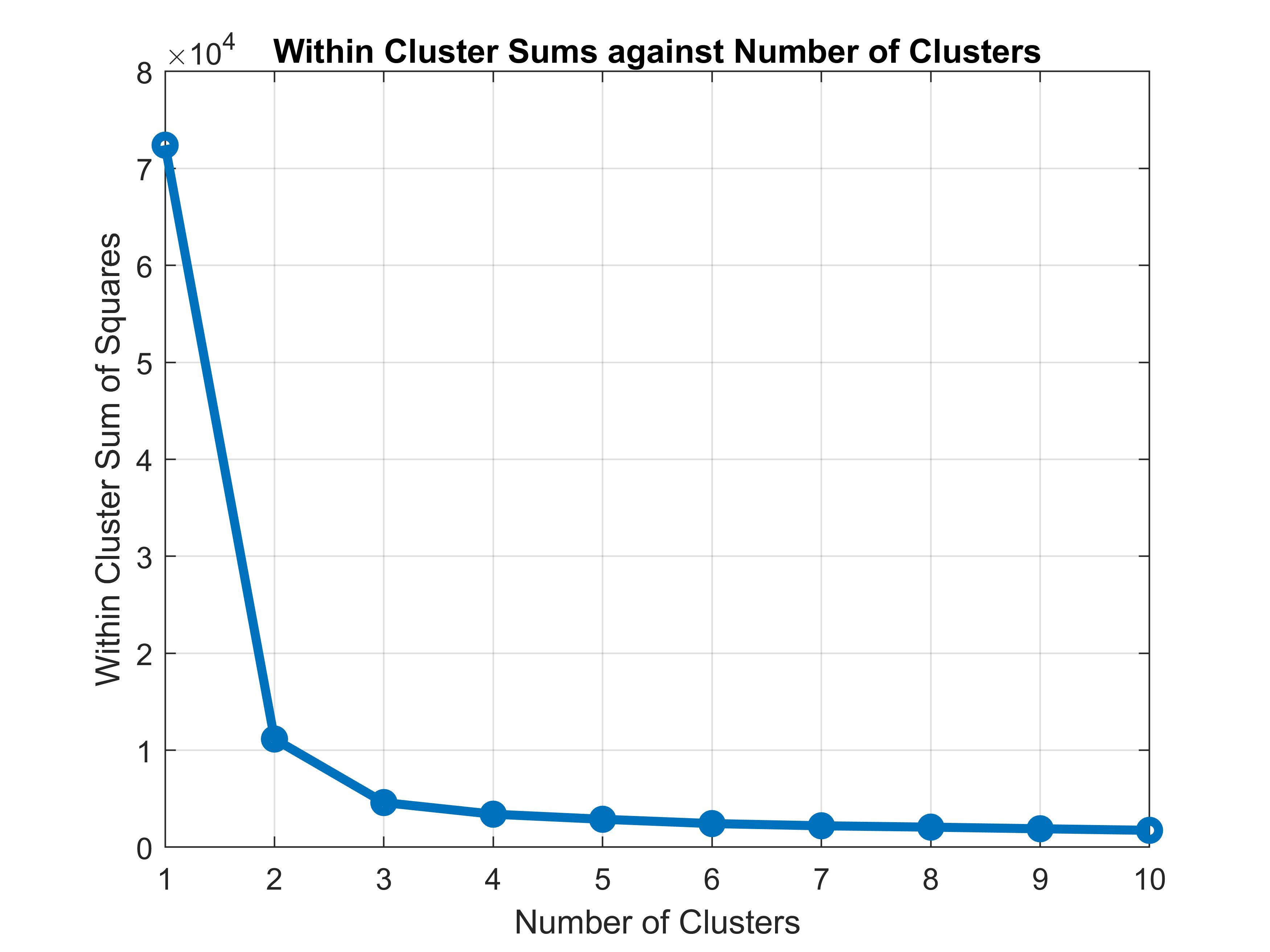}}%
\hfill 
\subcaptionbox{Mannar}{\includegraphics[width=0.25\textwidth]{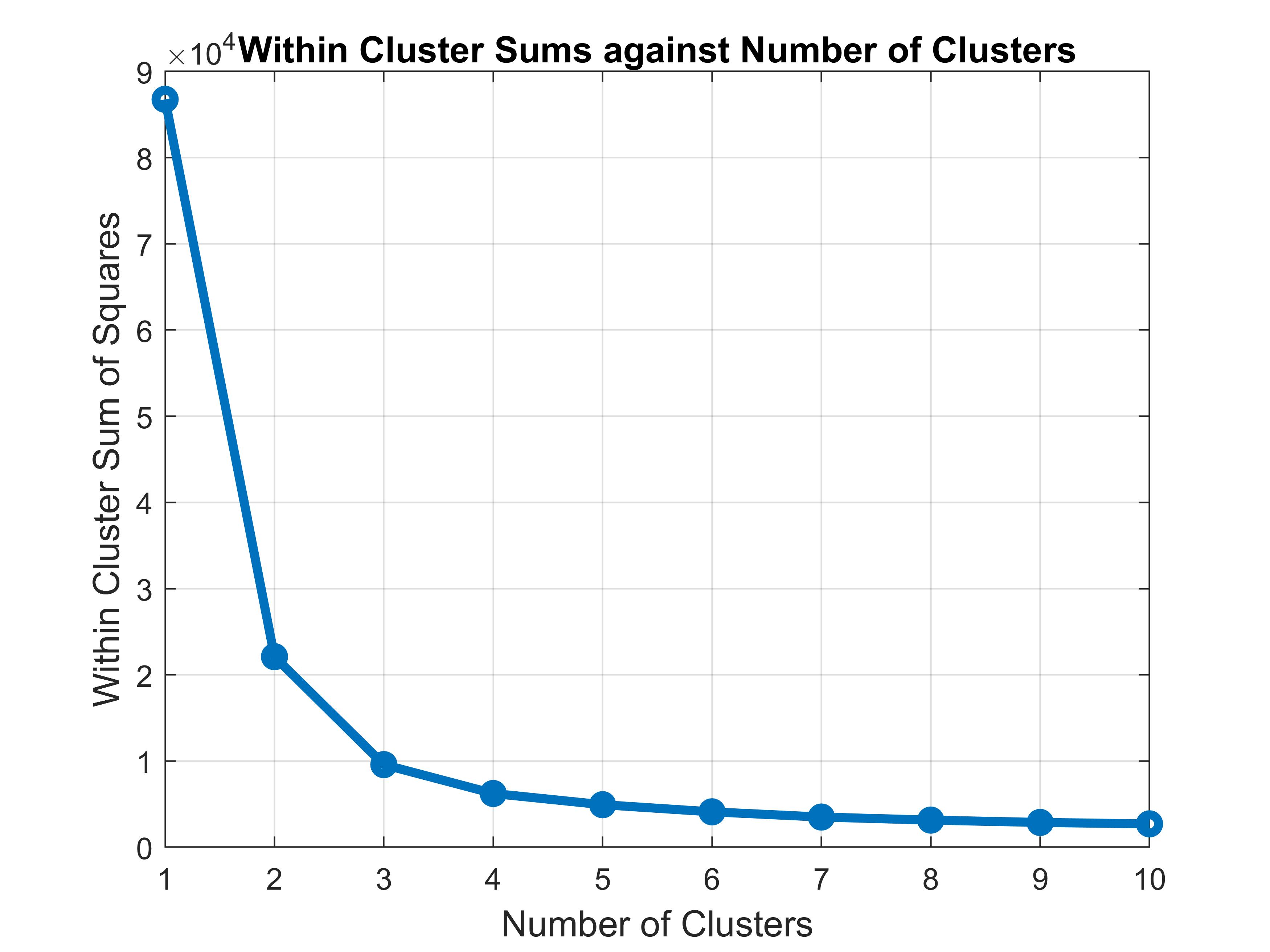}}%
\hfill 
\subcaptionbox{Giant's Tank}{\includegraphics[width=0.25\textwidth]{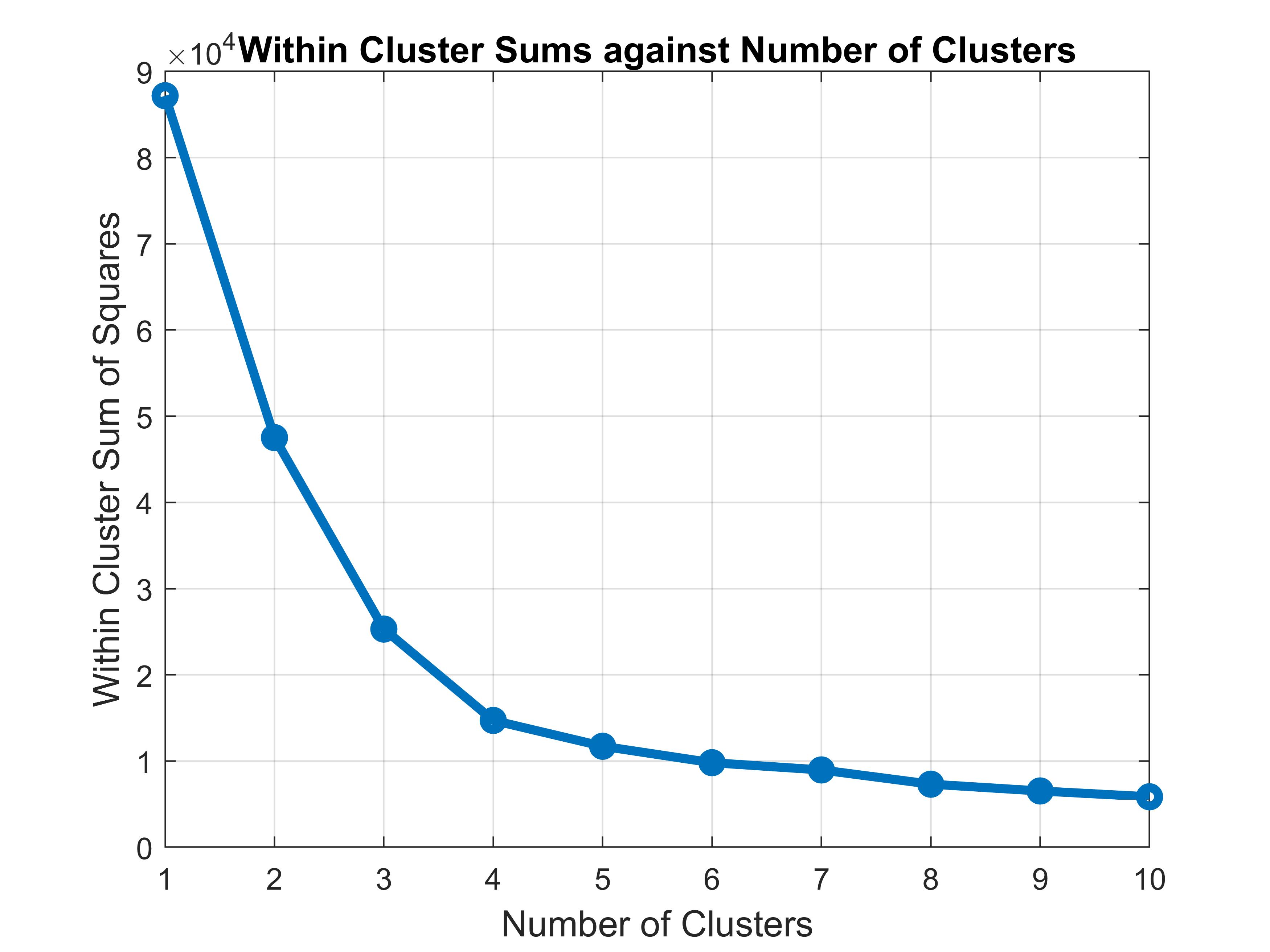}}%
\caption{Variation of the WCSS with the number of Clusters for each Region}
\label{figure: kmeans}
\end{figure}

As described in section \ref{subsection: pre-classification}, the image data of a particular terrain underwent iterative clustering by varying the number of clusters from one to ten. To determine the number of endmembers using the elbow method, the within-cluster sum of squares (WCSS) was calculated for each iteration, and a plot of WCSS vs. Number of Clusters was generated. Typically, the optimal number of clusters for the data is the number corresponding to the elbow point on the plot. The plots for the four terrains are illustrated in Fig. \ref{figure: kmeans}. After identifying the elbow points, the number of dominant endmembers in each location was determined and tabulated in Table \ref{tab:macroscopic endmembers}.

\begin{table}[h!]
    \centering
    \caption{Macroscopic Endmembers for the four Regions}
    \label{tab:macroscopic endmembers}
    \begin{tabular}{@{}ccc@{}}
        \toprule
        Geographical Location & No. of Endmembers & Identified Endmembers          \\ \midrule
        Jaffna                & 4                 & Water, Vegetation, Soil, Sand  \\
        Pulmoddai              & 3                 & Water, Soil, Vegetation        \\
        Mannar                & 3                 & Water, Soil, Vegetation        \\
        Murunkan              & 4                 & Water, Soil, Vegetation, Cloud \\ \bottomrule
    \end{tabular}
\end{table}

From Fig. \ref{figure: kmeans}, it can be seen that the WCSS value dropped significantly for Mannar, Pulmoddai and Jaffna when the number of clusters was set to two. In comparison, the WCSS of the Giant’s Tank does not drop as significantly. When considering the WCSS plots in general, a significant drop in the WCSS value demarcates the separation of a large cluster with a low variance. In this context, the above observation for Mannar, Pulmoddai and Jaffna results from the separation of highly abundant water pixels in the HSIs of those regions. On the contrary, in the Giant’s Tank’s HSI, the absence of such a large cluster of low variance could be attributed to the insignificant difference in the WCSS after forming two clusters. Therefore, one can get a sense of the presence of an abundant endmember with low variance in a given region by observing the WCSS plots.

\subsection{Identification and Isolation of Macroscopical Components}
\label{section: Identification and Isolation of Macroscopical Components}

\begin{figure}[H]
\centering
\subcaptionbox{Jaffna}{\includegraphics[width=0.25\textwidth]{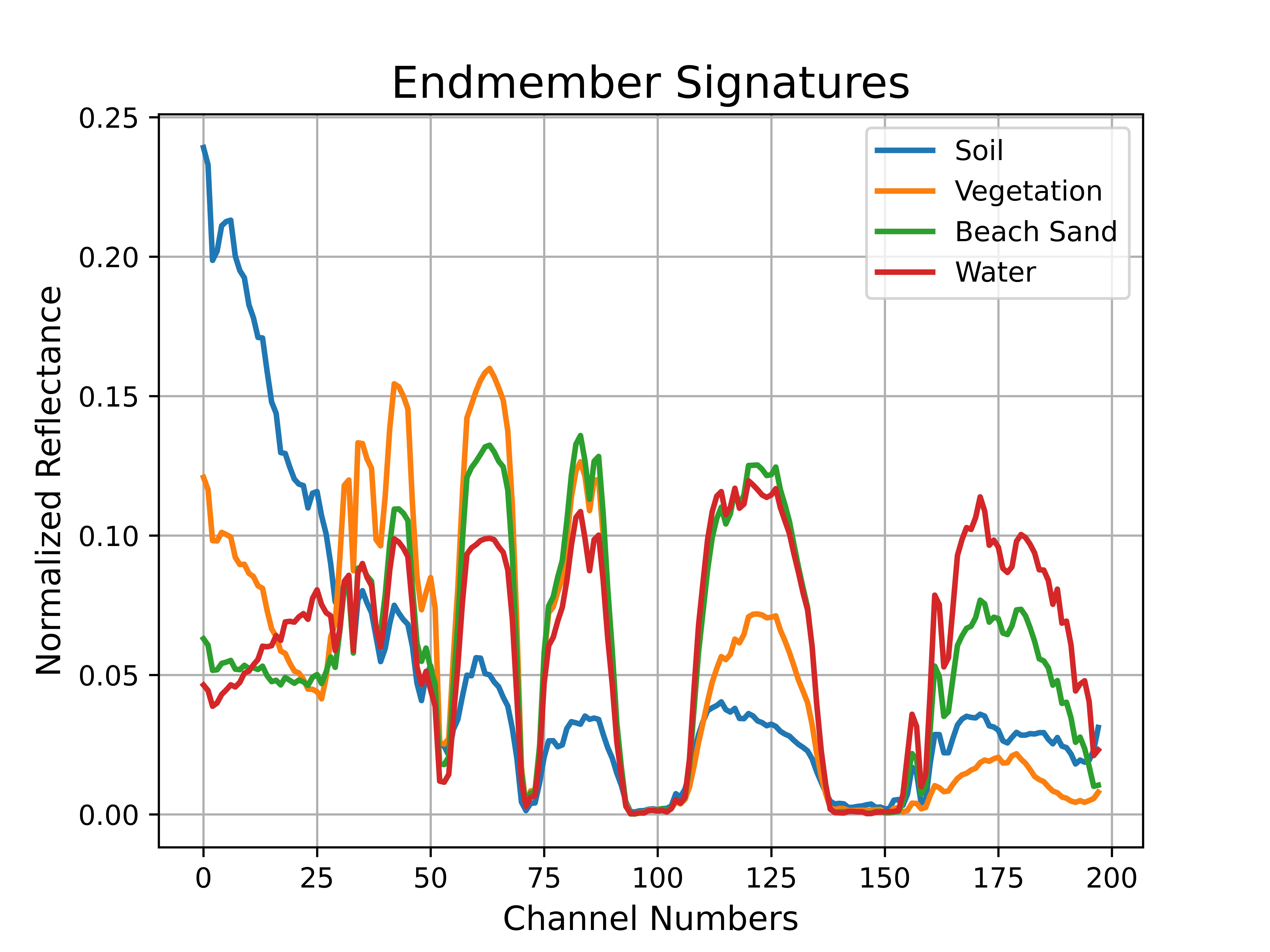}}%
\hfill 
\subcaptionbox{Pulmoddai}{\includegraphics[width=0.25\textwidth]{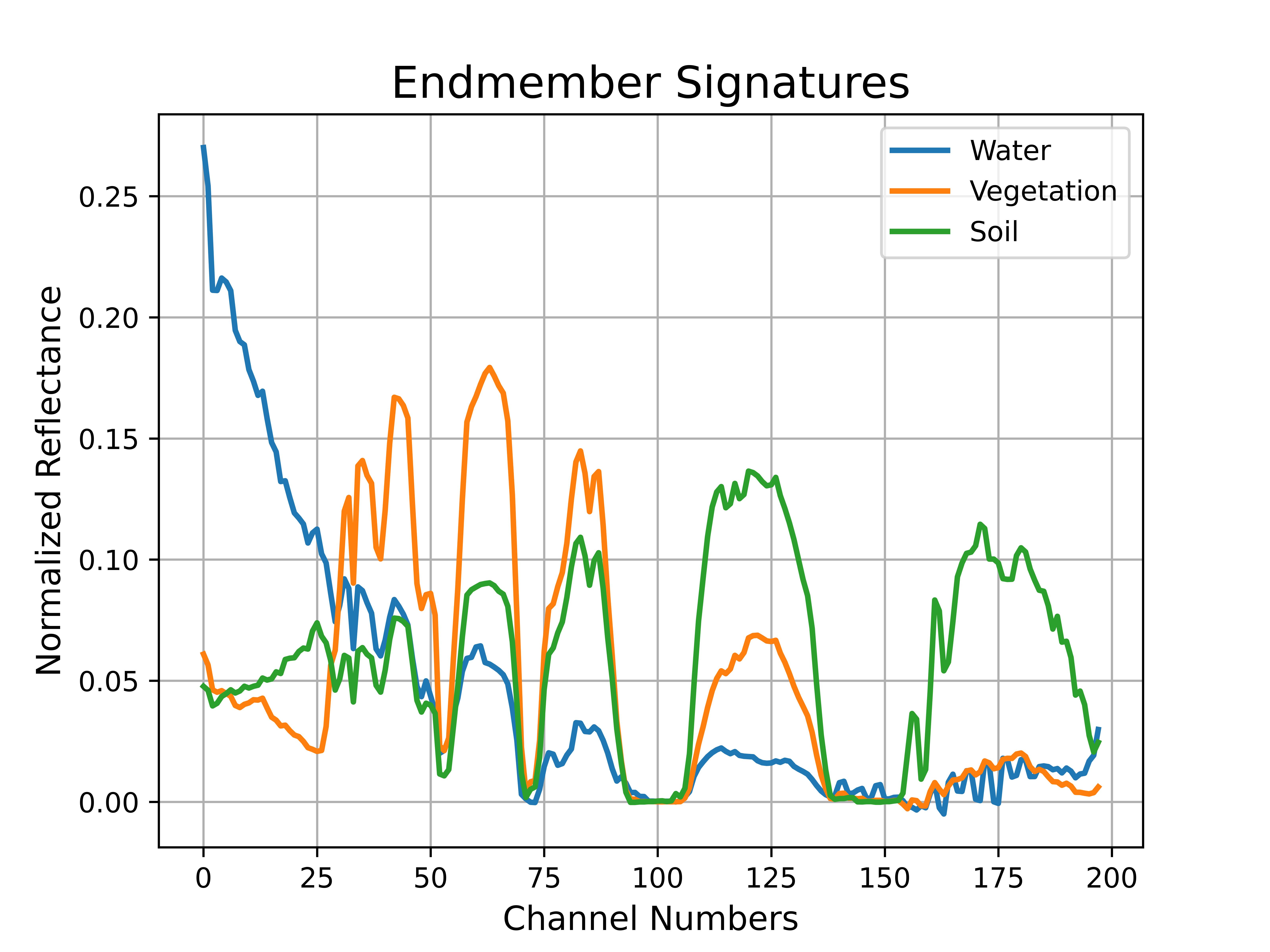}}%
\hfill 
\subcaptionbox{Mannar}{\includegraphics[width=0.25\textwidth]{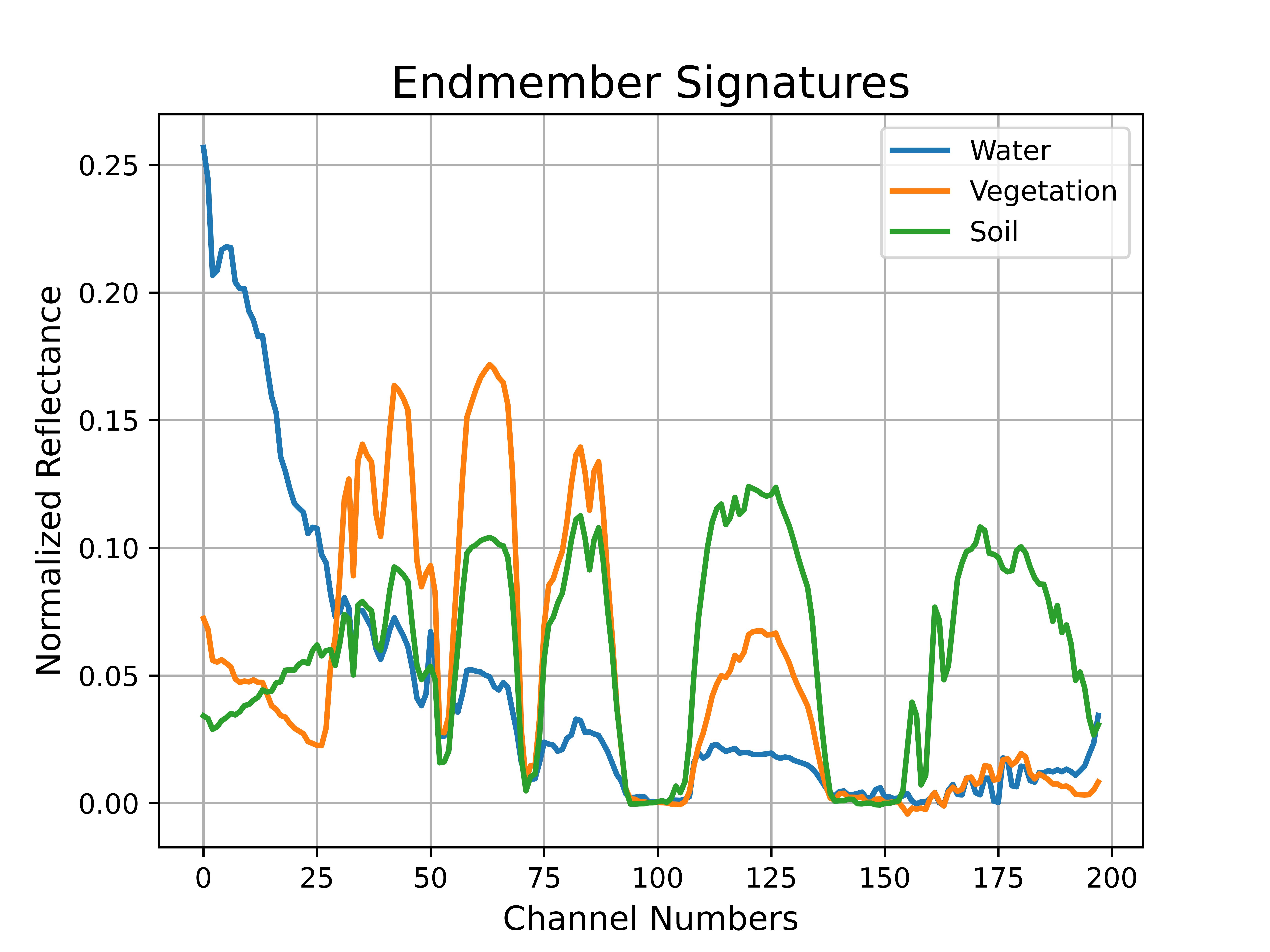}}%
\hfill 
\subcaptionbox{Giant's Tank}{\includegraphics[width=0.25\textwidth]{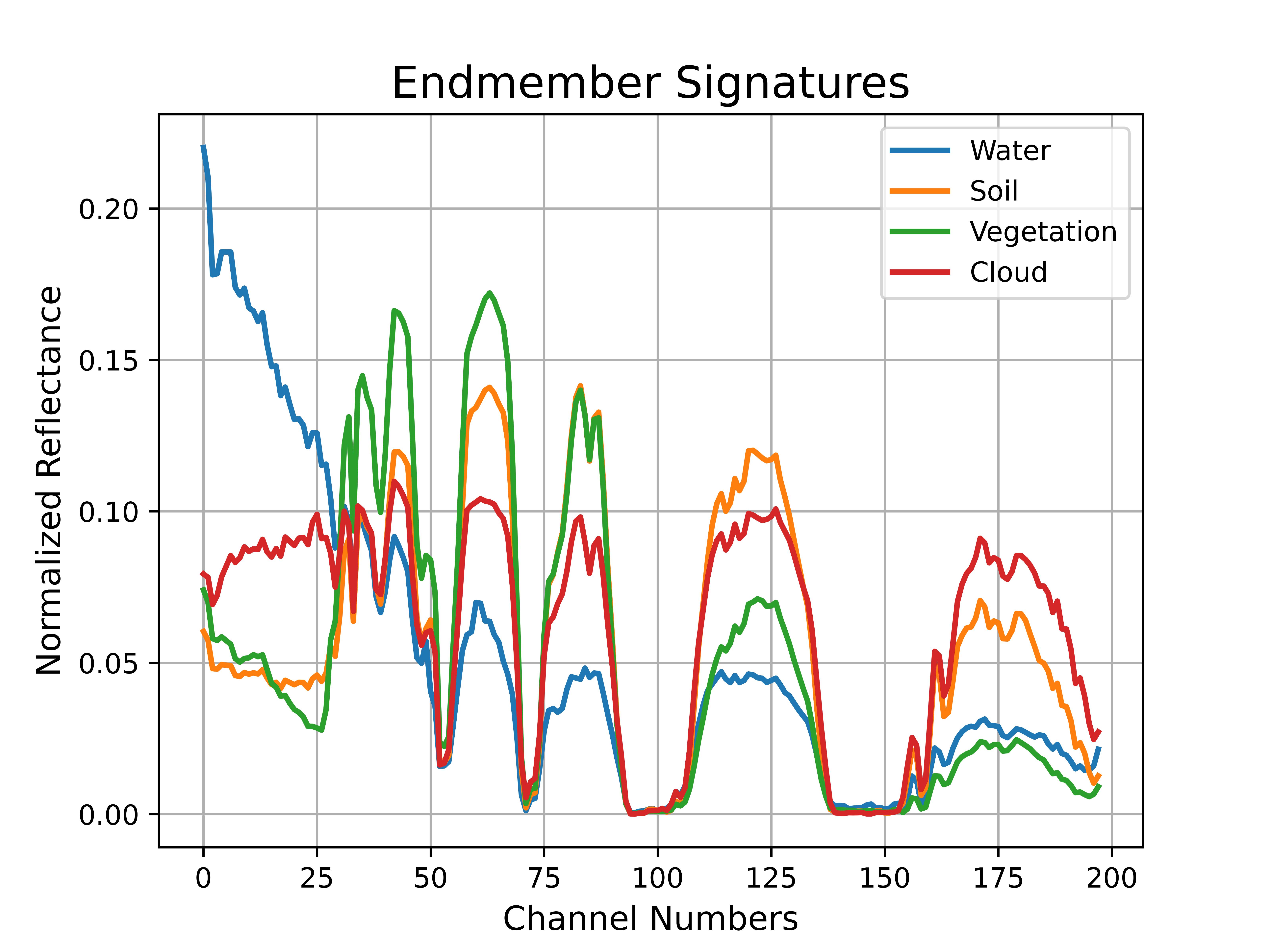}}%
\caption{The Macroscopic Endmember signatures extracted for the four locations considered}
\label{figure:macroscopic endmembers}
\end{figure}

With insights gained from the elbow method regarding the number of endmembers, VCA was employed by the authors to extract the dominant endmembers from each hyperspectral image (HSI). The extracted endmembers are visually depicted in Fig. \ref{figure:macroscopic endmembers}. In order to identify and isolate pixels corresponding to each candidate endmember, the authors utilized the similarity criteria mentioned in section \ref{subsection: pre-classification}. It is evident that the pre-classification stage successfully captured the land profile, effectively segmenting areas corresponding to different macroscopical components. Consequently, the identified endmembers were labelled with their respective macroscopical elements, as summarized in Table \ref{tab:macroscopic endmembers}.

The generated maps show that the algorithm has identified the regions corresponding to each endmember, allowing the isolation of soil pixels, which are used to perform further analysis for single-target mineral detection. While the proposed algorithm for the pre-classification block performed admirably, it is imperative to explore the adjustability and scalability of the proposed structure as some applications would require more refined extraction of end members and identification of the land cover of the particular endmember.

\begin{figure}[H]
\centering
\subcaptionbox{Jaffna}{\includegraphics[width=0.25\textwidth]{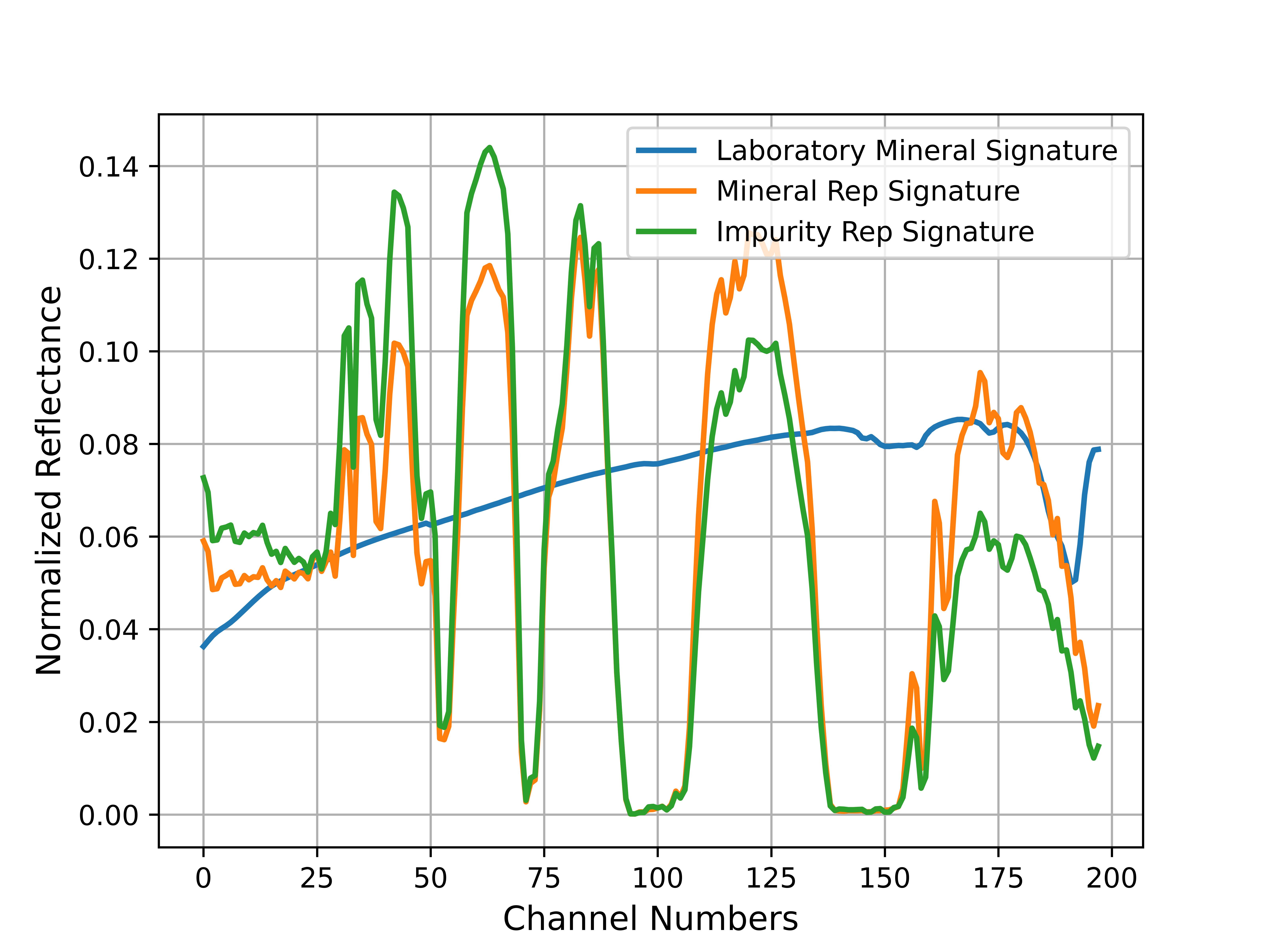}}%
\hfill 
\subcaptionbox{Pulmoddai}{\includegraphics[width=0.25\textwidth]{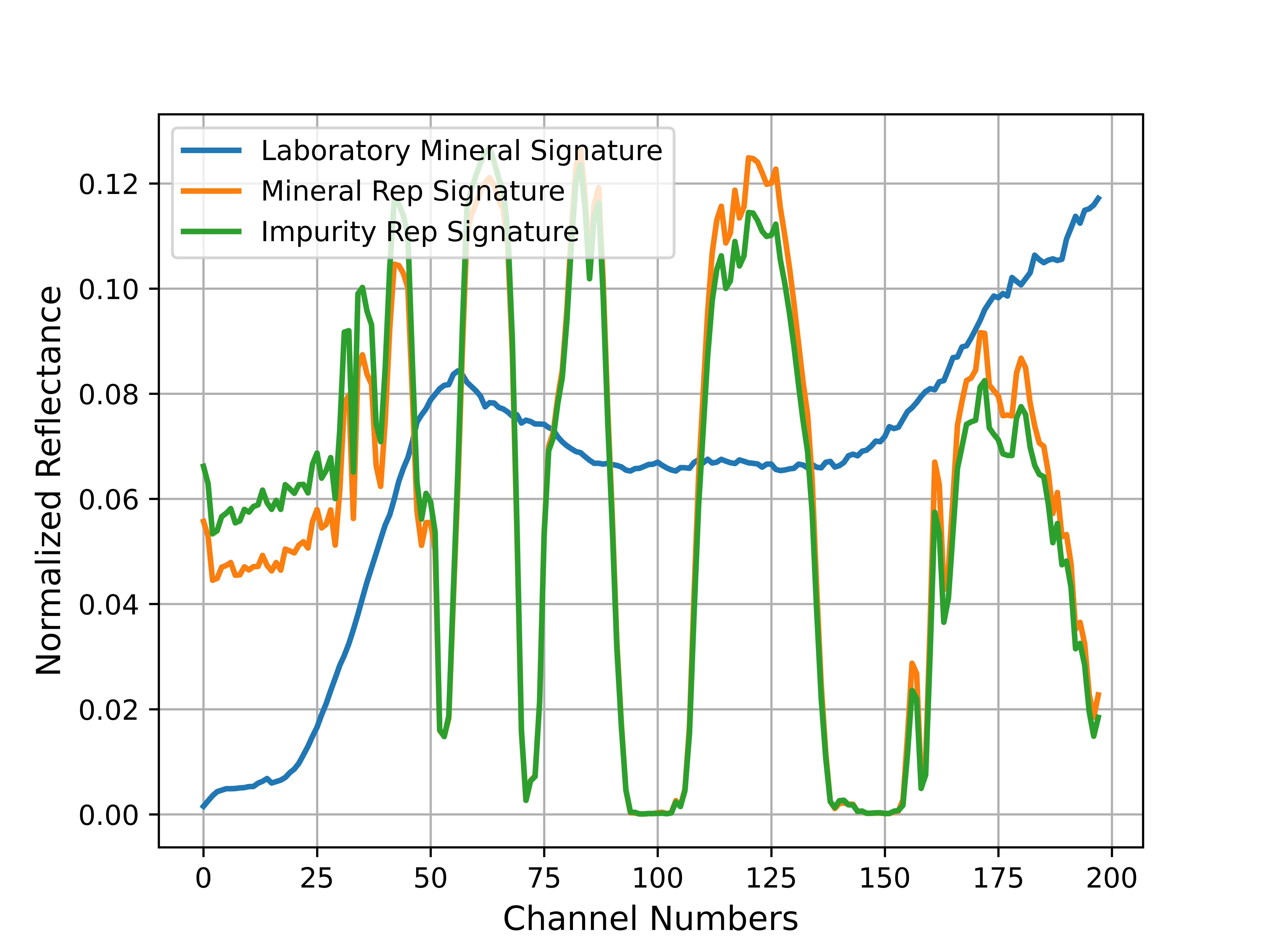}}%
\hfill 
\subcaptionbox{Mannar}{\includegraphics[width=0.25\textwidth]{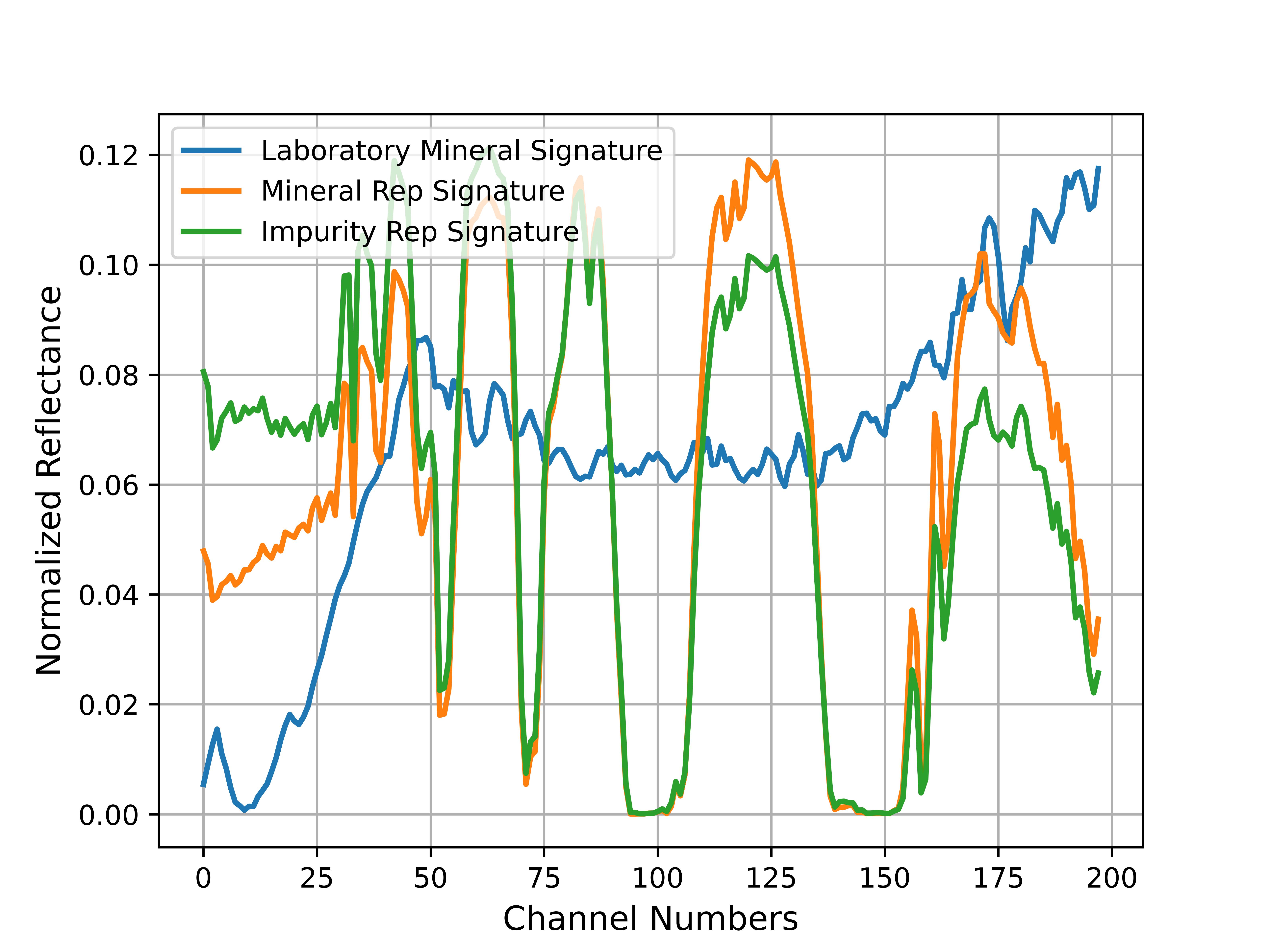}}%
\hfill 
\subcaptionbox{Giant's Tank}{\includegraphics[width=0.25\textwidth]{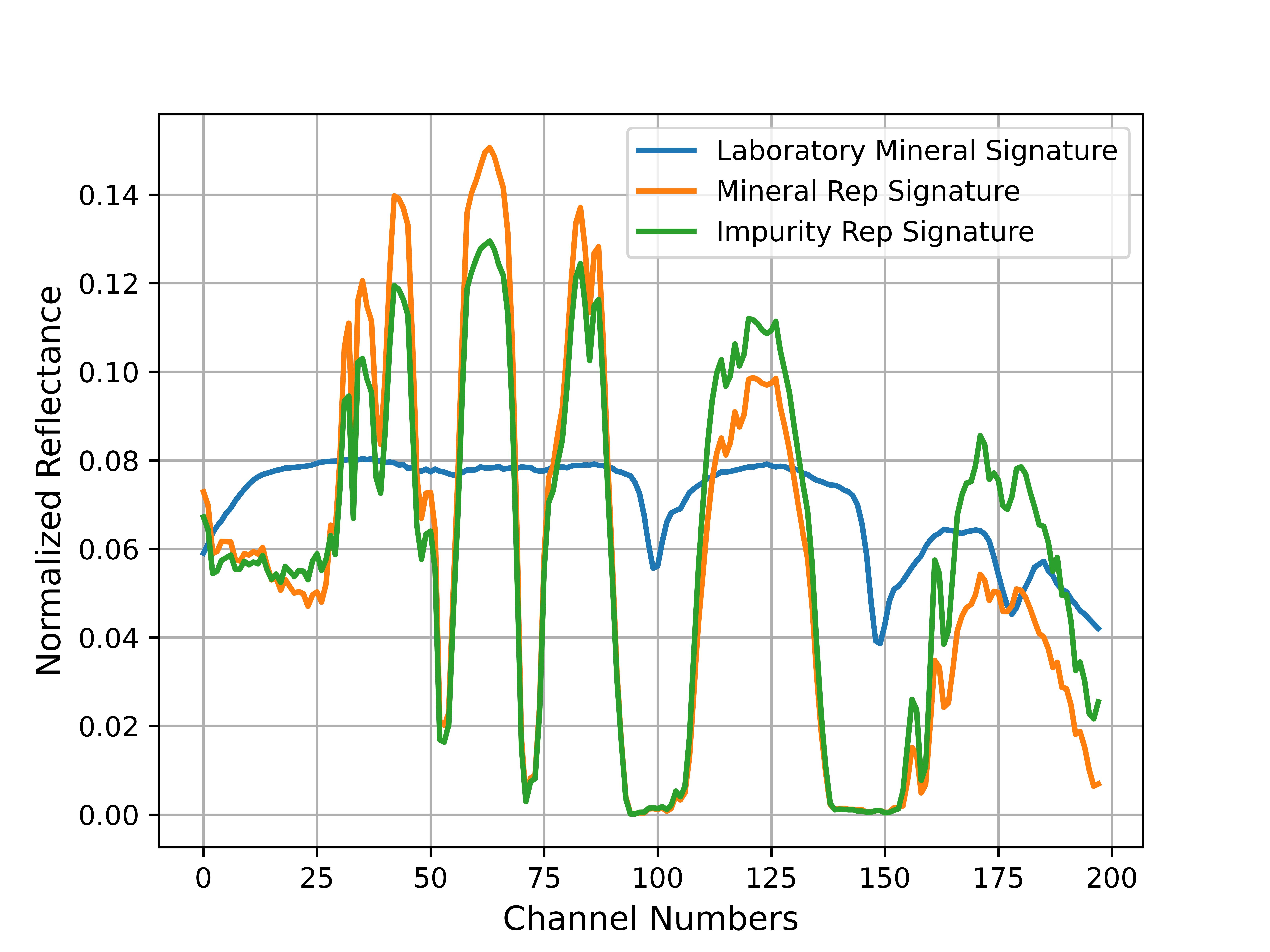}}%
\caption{The Mineral and Impurity representative signatures and the Laboratory Mineral signature for each location}
\label{figure:microscopic endmembers}
\end{figure}

\subsection{Correlation Analysis on Soil Pixels}
\label{Correlation Analysis on Soil Pixels}

Once the pixels belonging to the soil subclass were isolated using the methodology mentioned earlier, their correlation with laboratory mineral signatures obtained from the USGS Spectral Library was calculated. Per section \ref{subsection: remotely sensed data}, the extracted endmembers from the spectral library underwent several pre-processing steps to generate the signatures corresponding to Hyperion wavelengths. These pre-processed signatures provide a valuable reference for comparison and analysis in the subsequent stages of the study.

To determine the representative classes for each study, the first step involved extracting two signatures from the soil manifold. Subsequently, classification was performed using the algorithm outlined in section \ref{subsection: mineral-residual impurity classification}. In this classification process, the pixels were distinguished based on their correlation with the laboratory mineral signature. The range of correlation values for each study and the upper and lower bounds necessary for identifying the representative classes are summarised in Table \ref{tab:correlation analysis}. This table overviews the correlation thresholds for assigning pixels to the mineral or impurity representative classes.

\begin{table}[]
    \centering
    \caption{Results Generated through the Correlation Analysis}
    \label{tab:correlation analysis}
    \begin{tabular}{@{}cccc@{}}
        \toprule
        Site     & Range             & Lower Bound & Upper Bound \\ \midrule
        Jaffna   & -0.3808 to 0.0460 & -0.2518     & -0.0096     \\
        Pulmoddai & -0.5051 to 0.1886 & -0.0114     & 0.0300      \\
        Mannar   & -0.5015 to 0.2290 & -0.0925     & 0.1201      \\
        Murunkan & 0.4024 to 0.5816  & 0.4419      & 0.5531      \\ \bottomrule
    \end{tabular}
\end{table}

Based on the range of correlation values, it is evident that the laboratory signature of Montmorillonite has a positive correlation with pixel signatures belonging to a specific soil subclass in the area. This correlation is supported by the positive correlation between the laboratory signature and the two representative signatures extracted from the soil subclass. However, in the Jaffna terrain, the soil pixels are more likely to correlate negatively with Limestone's laboratory signature. Furthermore, the correlation range for soil pixels in Pulmoddai and Mannar, with respect to the Ilmenite reference signature, is quite similar. These findings align with the mineralogical background discussed in the corresponding section.

Once the subclasses have been identified according to section \ref{subsection: mineral-residual impurity classification}, the next step involves the replacement of the representative signatures by the mean of the pixel signature of each subclass. Fig. \ref{figure:microscopic endmembers} illustrates the laboratory mineral signature and the updated mineral and impurity representative signature for each mineral of interest.

\begin{figure}[]
\centering
\subcaptionbox{Jaffna}{\includegraphics[width=0.25\textwidth]{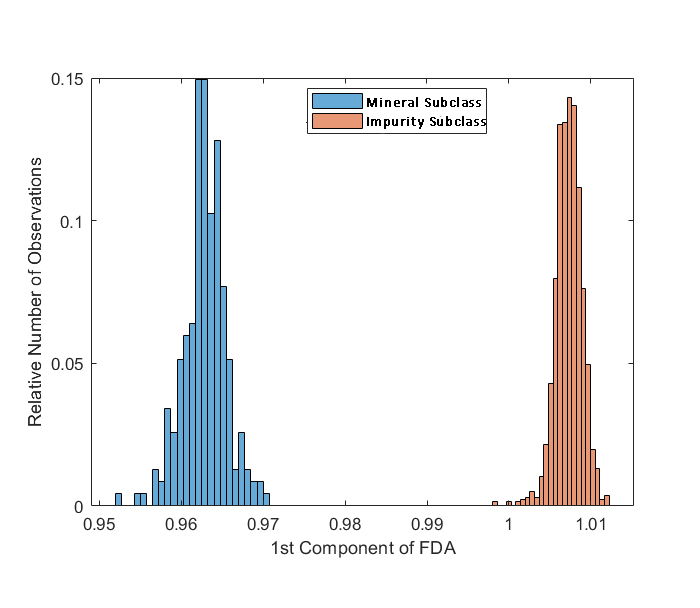}}%
\hfill 
\subcaptionbox{Mannar}{\includegraphics[width=0.25\textwidth]{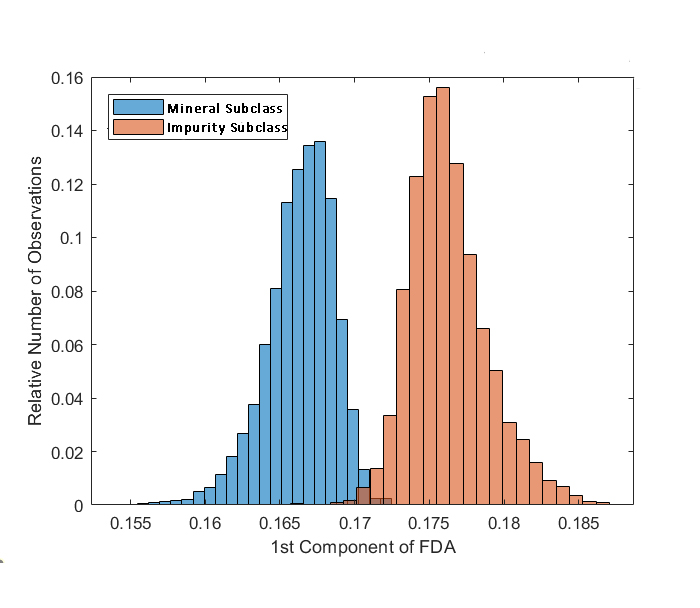}}%
\hfill 
\subcaptionbox{Giant's Tank}{\includegraphics[width=0.25\textwidth]{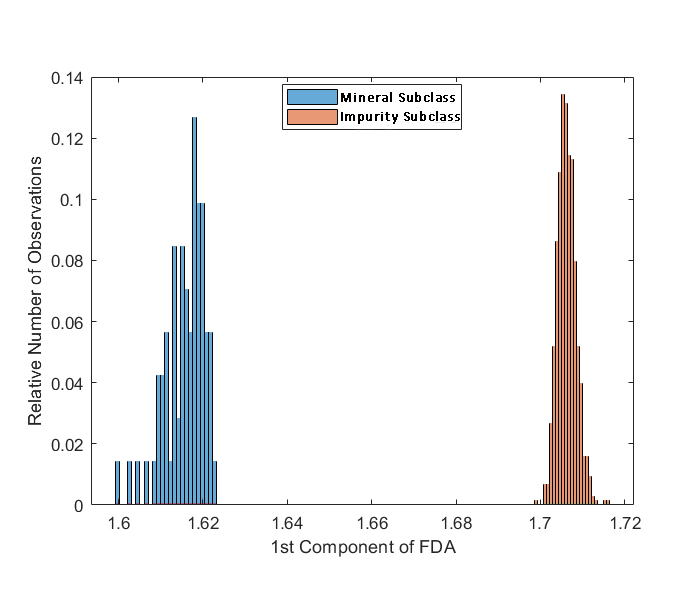}}%
\hfill 
\subcaptionbox{Pulmoddai}{\includegraphics[width=0.25\textwidth]{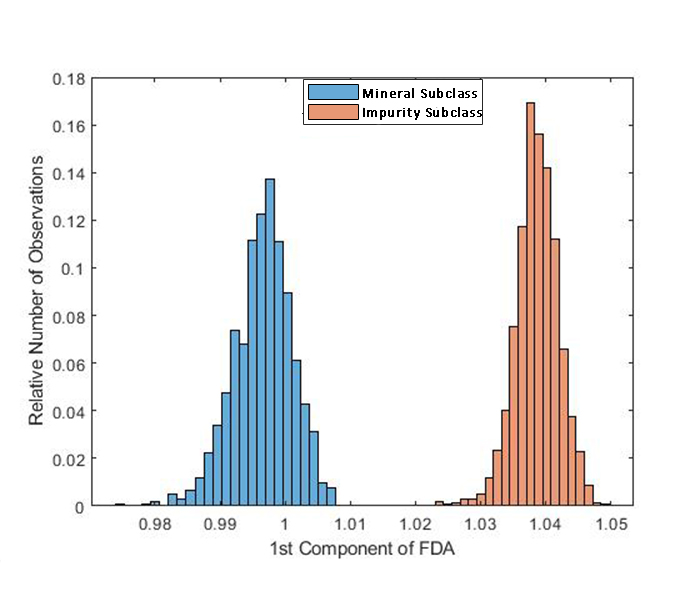}}%
\caption{Mineral and Impurity separaton in the reduced domain}
\label{fda plots}
\end{figure}

To develop the parameter relative availability, the eigendirection, which enhances the separation between the two subclasses, was computed using Fisher's Discriminant Analysis for each area of focus. Fig. \ref{fda plots} show the presence of the two subclasses in the reduced space. 

From the above figures, it can be seen that FDA has found an eigen-direction such that there is a good amount of separability between the two subclasses. 

It can be observed that the Montmorillonite pixels after being projected to the eigen-direction, show the largest relative separation (0.1) between the mean of the two subclasses. However, the signatures of the other three regions do not show separability (0.02 - Limestone, 0.01 - Mannar) compared to the Montmorillonite pixels. Moreover, it can be seen that the two clusters representing mineral representative and impurity representative pixels corresponding to Limestone and Pulmoddai show a reasonable separation in relation to the variance of the clusters. On the contrary, Fig. \ref{fda plots} corresponding to Mannar depicts a slight overlap between the two representative clusters. This might be because the site contains a low concentration of the mineral. Less mineral presence in a given pixel increases the presence of the impurities, pushing the mineral subclass closer to the impurity subclass. This claim is proven by the XRD validation of the samples collected from the Mannar region. The samples collected from the region show less mineral presence, which has caused the two subclasses to be close and slightly overlapping.

\subsection{Abundance Generation and Laboratory Test Results}

In order to generate the abundance map for a particular mineral, the authors derive the parameter relative availability in the reduced space.  Depending on the mean representative signatures derived depending on the relative availability measure  for each class, the authors calculate the abundance for each soil pixel, assuming a linear mixture model for each pixel. In this section the abundances computed through the algorithm is compared with the laboratory test results obtained for each mineral for each region.

\subsubsection{Mannar - Ilmenite} 

\begin{figure}[]
\centering
\subcaptionbox{}{\includegraphics[width=0.3\textwidth]{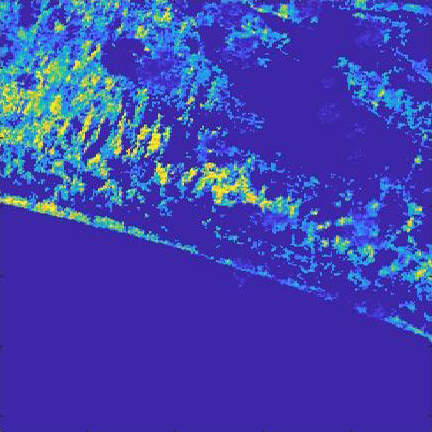}}%
\hfill 
\subcaptionbox{}{\includegraphics[width=0.3\textwidth]{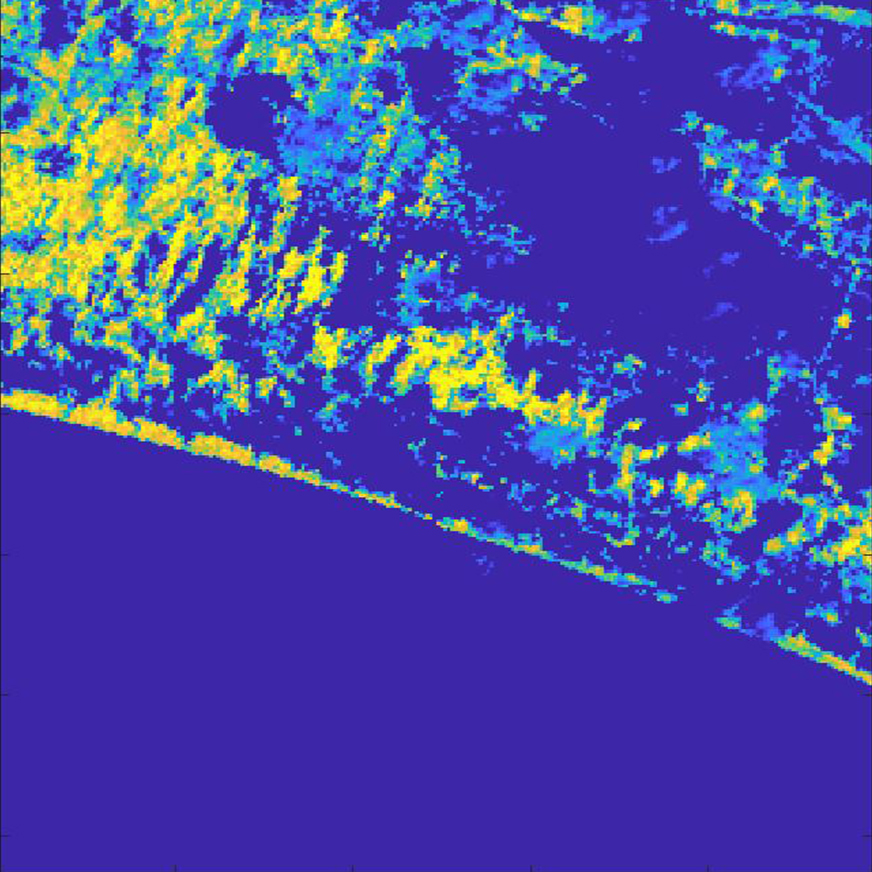}}%
\hfill 
\subcaptionbox{}{\includegraphics[width=0.3\textwidth]{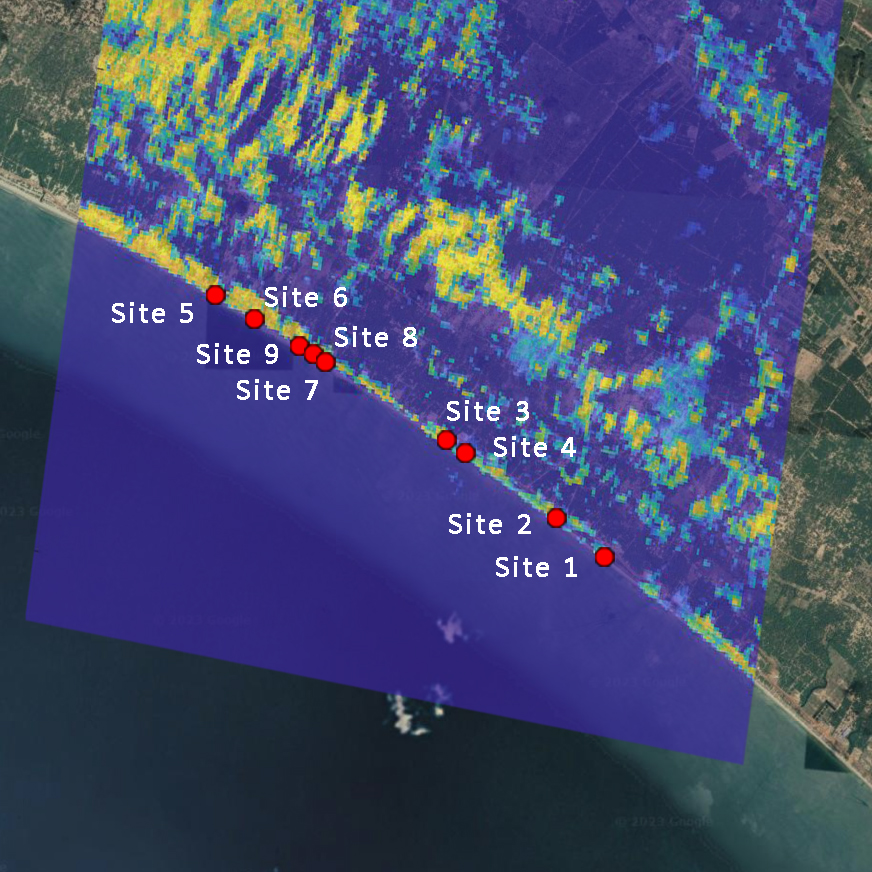}}%
\caption{(a) Relative availability Map (b) Abundance map generated after NNLS (Alpha values generated after NNLS) (c) Generated mineral map overlapped with site}
\label{figure: mannar relative availability}
\end{figure}

Mannar is known to be the fourth largest Ilmenite deposit in the world. The soil of Mannar island is considered to be mineral rich soil containing Ilmenite, leucoxene, zirconium, rutile, titanium oxide, granite, sillimanite or orthoclase. For the validation of the proposed algorithm, Mannar is chosen as a region of interest for finding possible Ilmenite deposits. Fig. \ref{figure: mannar relative availability} (a) illustrates the relative availability values generated as an intermediate result while Fig. \ref{figure: mannar relative availability} (b) illustrates the abundances predicted via the linear mixture model. In order to prove the accuracy and applicability of the proposed algorithm, nine sites were selected on the southern coast of the Mannar island based on the $\alpha$ values Fig. \ref{figure: mannar relative availability} (c). As mentioned in section \ref{magnetic separation}, due to the fact that the recommended grain size for magnetic separation was not applicable for the Ilmenite in Mannar, XRD tests were carried out to measure the Ilmenite availability. The results obtained through the XRD tests conducted on the samples of the ten sites are tabulated in table \ref{tab:mannar test results}.

\begin{table}[]
    \centering
    \caption{XRD Test Results, Relative Availability and Abundance Estimation after Non-negative Least Squares (NNLS) for the Sites at Mannar}
    \begin{tabular}{@{}cccc@{}}
        \toprule
        Site                       & XRD Test Results            & Relative Availability      & Abundance after NNLS Estimate \\ \midrule
        Site 1 & 0.29\% & 0.4649         & 0.4679 \\
        Site 2 & 0.46\% & 0.445          & 0.4258 \\
        Site 3 & 0.09\% & 0.4563         & 0.4958 \\
        Site 4 & 0.21\% & 0.3743         & 0.5863 \\
        Site 5 & 0.36\% & 0.705          & 0.7982 \\
        Site 6 & 0.39\% & 0.5619         & 0.6587 \\
        Site 7 & 0.49\% & 0.7275         & 0.8513 \\
        Site 8 & 0.36\% & 0.668          & 0.8134 \\
        Site 9 & 0.35\% & 0.6358         & 0.6971  \\ \midrule
    \end{tabular}
    \label{tab:mannar test results}
\end{table}

The results range from the lowest value of 0.09\% to the highest value of 0.49\%. From these results, it is apparent that the percentage Ilmenite availability is low for the selected sites. As illustrated in Fig. \ref{fda plots} of section \ref{Correlation Analysis on Soil Pixels}, the algorithm shows a slight overlap between the mineral and impurity representative classes. From the XRD results it can be confirmed that this overlap might be caused as a result of misclassification of pixels due to low mineral concentration present in the site, as mentioned in section \ref{Correlation Analysis on Soil Pixels}. The intermediary results generated as the relative availability show traces of Ilmenite along the coast and as strips going from the coast towards the inland. Furthermore, it can be seen that there is a good correspondence between the XRD test results and the relative availability. However, when calculating the correlation coefficient between them, the coefficient resulted in a value of 0.5640. This value could be due to the fact that the pixels considered for the sites may contain a high portion of water. Thus, it could have led the predictions to not be highly correlated with the XRD values. The linear mixture model has further enhanced the abundances from the previous step, showing a high availability in the regions mentioned before. 

The algorithm has managed to retrieve most of  the sites with comparatively high Ilmenite percentage through the abundance values. However, it can be seen that  the algorithm has performed poorly when assigning abundance value for the site with lowest Ilmenite percentage. This might be caused due to the misclassification of impurity pixels as mineral representatives due to the overlapping of the two classes. Throughout most of the sites of interest, the algorithm has managed to preserve the positively correlated pattern between the laboratory XRD results with the abundances ($\alpha$) generated through the algorithm. Even Though the percentage mineral availability is low for the selected region of Mannar, the algorithm performs satisfactorily.

\subsubsection{Pulmoddai}

For the case of Pulmoddai, samples for the validation of the proposed algorithm were collected from eight different sites based on the predictions of the algorithm. Some sites were towards the coastal region while others were scattered along the distributors of Yan oya, which flows into the sea from Pulmoddai at Pangurugaswewa in Trincomalee district. Specifically, sites 1, 2, 3 and 4 were selected from the coastal region while sites 5, 6, 7 and 8 were collected from the southern bank of Yan oya.

As mentioned in section \ref{magnetic separation} magnetic separation method was followed to identify the Ilmenite availability of the collected samples. The results obtained for each site are tabulated in table \ref{tab: pulmoddai test results}. It can be seen that the samples collected from coastal sites 3 and 4 yield the largest Ilmenite percentage. However, the sites along the banks of Yan oya also provide a comparatively high percentage availability for Ilmenite. Comparing the results of the proposed algorithm with the magnetic separation outcomes it is evident that the algorithm has performed effectively. The results obtained for relative availability and alpha have managed to preserve the patterns observed in the magnetic separation results to some extent. The correlation coefficient between the magnetic separation results with that of the predicted values was 0.8115. This further corroborates the capability of the algorithm at predicting the sites' mineral availability. The algorithm suggests the presence of Ilmenite with a relatively high abundance ($\alpha$) for the selected coastal sites 3 and 4 which is well supported by laboratory test results. Moreover the algorithmic suggestion for abundance shows a positive correlation to some degree with the real Ilmenite percentages in  the soil. It assures the capability of the algorithm to identify possible sites for surveying of Ilmenite. In addition the positive correlation shown hints a capability to enhance the algorithm further to predict the gradient of mineral concentration for Ilmenite in a given terrain allowing the ability to develop contours mapping for mineral availability.

\begin{table}[]
    \centering
    \caption{Magnetic Separation Test Results, Relative Availability and Abundance Estimation after NNLS for the Sites at Pulmoddai}
    \begin{tabular}{@{}cccc@{}}
        \toprule
        Site   & Magnetic Separation  & Relative Availability & Abundance after \\ 
              & Test Results          &                       & NNLS Estimate \\
        \midrule
        Site 1 & 0.59\% & 0.3670                & 0.3670                       \\
        Site 2 & 1.53\% & 0.3021                & 0.3954                       \\
        Site 3 & 12.27\%& 0.9893                & 0.9607                       \\
        Site 4 & 10.73\%& 0.4922                & 0.4922                       \\
        Site 5 & 1.95\% & 0.2917                & 0.3271                       \\
        Site 6 & 3.80\% & 0.4912                & 0.4565                       \\
        Site 7 & 5.70\% & 0.687                 & 0.687                        \\
        Site 8 & 8.17\% & 0.7694                & 0.7911                      \\ \midrule
    \end{tabular}
    \label{tab: pulmoddai test results}
\end{table}

For a mineral sand deposit to be considered suitable for mineral extraction depends on the heavy mineral grade (HM grade). Typically HM grade ranging from 0.5\% to 20\% and above is considered as suitable for the above purpose. In this light, all the samples which were collected  from Pulmoddai based on the results generated through the algorithm proposed,  warrants to be candidates for Ilmenite extraction. 

Previous studies and industrial resources indicate that Ilmenite in Pulmoddai region is mostly confined to the coastal region.However,  it can be seen that the algorithm has predicted there are probable deposits towards the inland. Importantly, the deposits predicted along the distributors of Yan oya supports the geological claim that rivers in Sri Lanka carry massive loads of heavy minerals over time to the coasts and are redistributed by the sea currents. It is further validated through the Ilmenite concentration found from the magnetic separation method. The predictions on the inland Ilmenite deposits in the Pulmoddai area , allows the opportunity for one to expedite the sites suggested by the algorithm in the future.

\subsubsection{Jaffna}

\begin{figure*}[]
    \centering
    
    \begin{subfigure}[t]{0.2\textwidth}
        \centering
        \includegraphics[width=\textwidth]{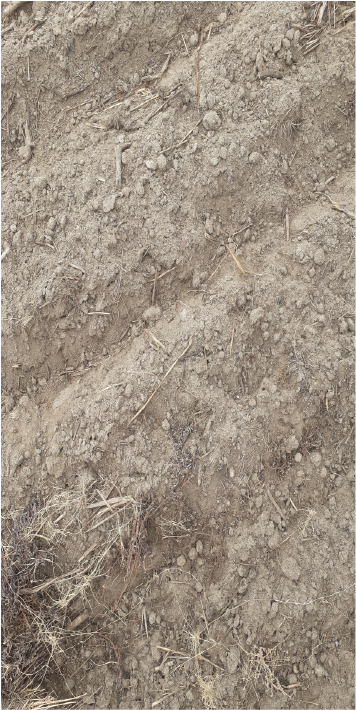}
        \caption{site-1}
        \label{figure: site-1}
    \end{subfigure}
    ~
    \begin{subfigure}[t]{0.2\textwidth}
        \centering
        \includegraphics[width=\textwidth]{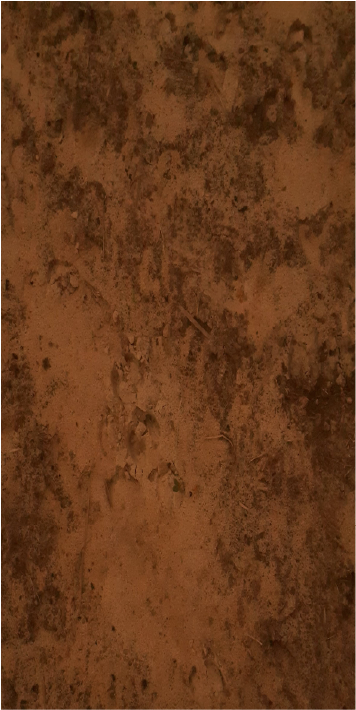}
        \caption{site-3}
        \label{figure: site-3}
    \end{subfigure}
    ~
    \begin{subfigure}[t]{0.2\textwidth}
        \centering
        \includegraphics[width=\textwidth]{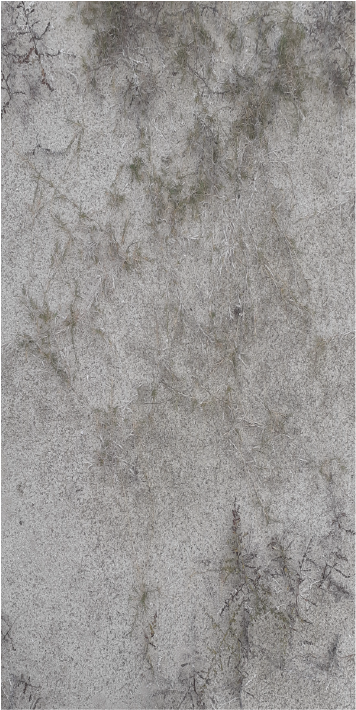}
        \caption{site-2}
        \label{figure: site-2}
    \end{subfigure}
    ~
    \begin{subfigure}[t]{0.2\textwidth}
        \centering
        \includegraphics[width=\textwidth]{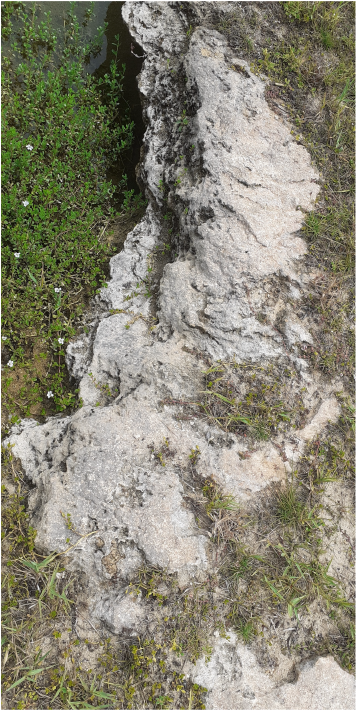}
        \caption{site-4}
        \label{figure: site-4}
    \end{subfigure}

    \caption{In-situ soil characteristic observations}
    \label{figure: soil characteristic visuals}
\end{figure*}

\begin{figure}[h!]
\centering
\subcaptionbox{}{\includegraphics[width=0.3\textwidth]{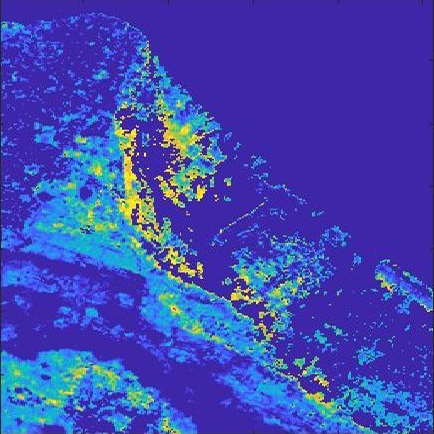}}%
\hfill 
\subcaptionbox{}{\includegraphics[width=0.3\textwidth]{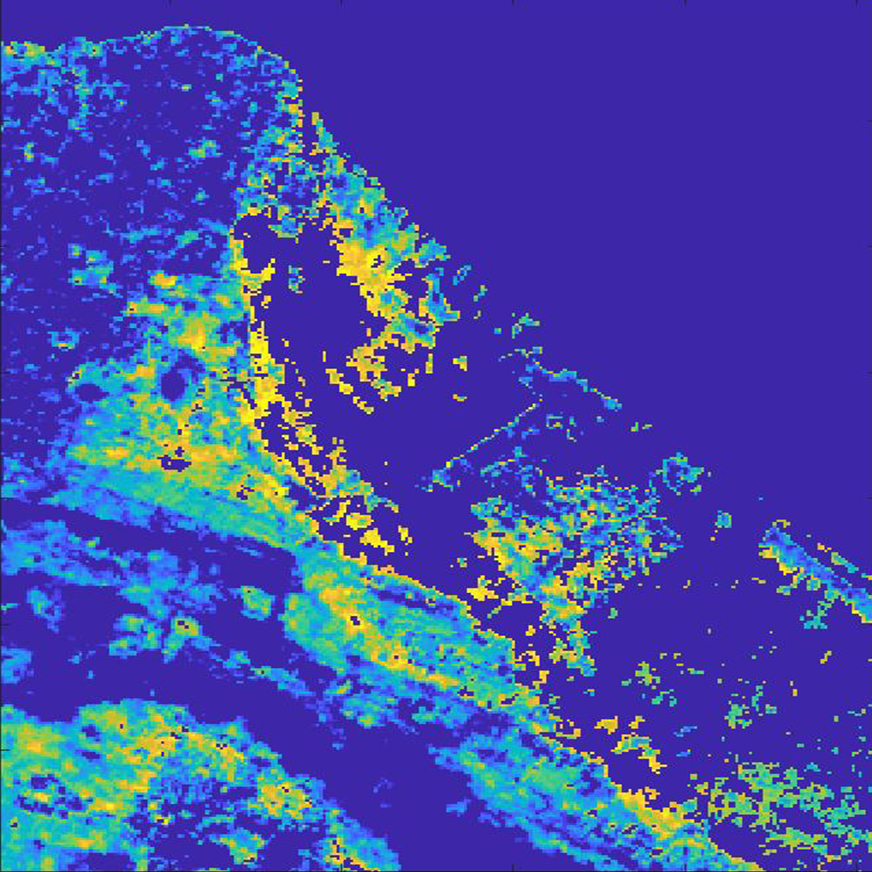}}%
\hfill 
\subcaptionbox{}{\includegraphics[width=0.3\textwidth]{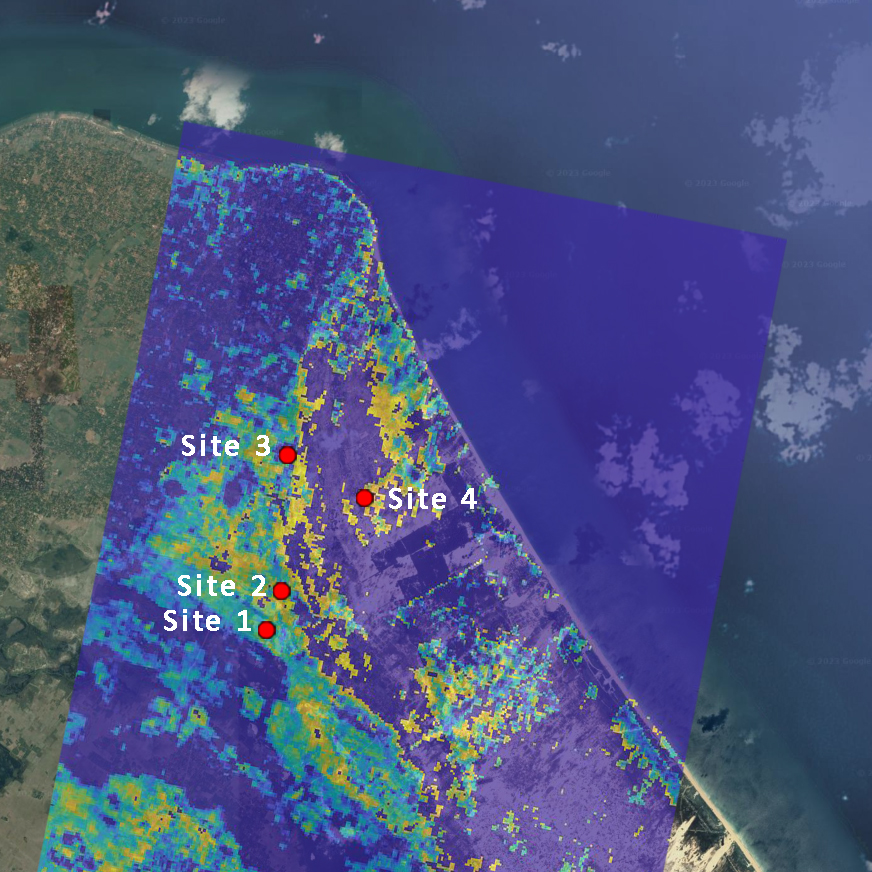}}%
\caption{(a) Relative availability Map (b) Abundance map generated after NNLS (Alpha values generated after NNLS) (c) Generated mineral map overlapped with sites of Jaffna}
\label{figure: jaffna relative availability}
\end{figure}

The provided Fig.\ref{figure: jaffna relative availability} illustrates the soil characteristics and surface texture of four survey sites. To begin with, site 1 (Fig.\ref{figure: site-1}) was characterised by sparse vegetation and had dry and compact soil, without any visible signs of Limestone in the vicinity. Moving on to site 3 (Fig.\ref{figure: site-3}), there was no apparent presence of surface Limestone, and the reddish-brown soil colour differed from the other three sites. However, it's worth noting that Site-3 was situated in a populated area and had experienced ongoing human activity, which may have led to abrupt changes in soil composition, differing from what would be expected based on historical geological conditions.In contrast, site 2 (Fig.\ref{figure: site-2}) featured moist, dark brown soil and exhibited seashell fragments scattered on the surface. This area was identified as a drained lagoon, and it was large enough to be discerned through satellite imagery. The widespread presence of seashell remnants throughout the lagoon area suggested their deposition due to tides, serving as evidence for the existence of Limestone on the surface.

\begin{table}[]
    \centering
    \caption{XRD Test Results, Relative Availability and Abundance Estimation after NNLS for the Sites at Jaffna}
    \begin{tabular}{@{}cccc@{}}
    \toprule
    Site   & XRD Test Results & Relative Availability & Abundance after NNLS Estimate \\ \midrule
    Site 1 & 1.42\%           & 0.0984                & 0.1269                       \\
    Site 2 & 24.24\%          & 0.3234                & 0.5874                       \\
    Site 3 & 4.88\%           & 0.1663                & 0.1022                       \\
    Site 4 & 48.87\%          & 0.7688                & 0.7652                       \\ \bottomrule
    \end{tabular}
    \label{tab: jaffna test results}
\end{table}

Lastly, site 4 (Fig.\ref{figure: site-4}) had a desiccated soil texture and displayed shell fragments, providing evidence of Limestone in the soil. Furthermore, during the survey, medium-sized Limestone fragments were discovered in the vicinity of site 4. This particular location was the closest to the shoreline, and it is plausible that these sediments were deposited there over time with the assistance of sea currents during the formation of the Jaffna peninsula.

To buttress the observations from these sites, the amount of Limestone available according to the XRD test is provided in Table \ref{tab: jaffna test results}. As tabulated in Table \ref{tab: jaffna test results}, the proposed method had made an accurate prediction of the Limestone presence in site 2 and site 4 with higher $\alpha$ values while returning low $\alpha$ values for site 1 and site 3 that implies less chance of Limestone presence, which is in agreement with the observations made by the validation field survey via XRD tests. Additionally, the correlation coefficient between the XRD values and the relative availability values resulted in a value of 0.9853. This further confirms the fact that algorithm has been able to accurately predict the availability in these sites. Furthermore, the relative availability and the  alpha values calculated for the four sites show a good correspondence with the XRD test results. Based on the results it is evident that the algorithm performs admirably across sites with different soil characteristics to find feasible survey sites for Limestone availability.

\subsubsection{Giant's Tank}

The availability of Montmorillonite in the Giant’s Tank and Murunkan regions in Sri Lanka has been confirmed by earlier research \citep{wanasinghe2017extraction}; \citep{herath1973industrial}. Montmorillonite, being a clay mineral, is abundant in arid regions, often found in tanks constructed for cultivation purposes. Geomorphologically, Murunkan primarily features flat terrain with numerous shallow water bodies nearby. These water bodies periodically dry up during extended droughts but are replenished mainly by rainwater. The clay-rich composition of the soil contributes to a high surface water retention capacity due to its low infiltration rate. Montmorillonite tends to accumulate at the bottom of these tanks through sedimentation and is also commonly found in the regions surrounding the tanks and the fields irrigated using water from these reservoirs.

This study specifically focuses on the region containing the Giant’s Tank, which is a prominent irrigation tank located in Northern Sri Lanka. This region is characterised by abundant wetlands and paddy fields, likely containing traces of Montmorillonite.

\begin{table}[h]
    \centering
    \caption{XRD Test Results, Relative Availability, and Abundance Estimation after NNLS for the Sites at Giant's Tank}
    \begin{tabular}{@{}cccc@{}}
    \toprule
    Site & \multicolumn{1}{c}{XRD Test Results} & Relative Availability & Abundance after NNLS Estimate \\ \midrule
    Site 1        & 4.40\%                                         & 0.5479                          & 0.6331                                  \\
    Site 2        & 1.20\%                                         & 0.5192                          & 0.5993                                  \\
    Site 3        & 32\%                                           & 0.8199                          & 0.9976                                  \\
    Site 4        & 2.10\%                                         & 0.5252                          & 0.6234                                  \\
    Site 5        & 7.10\%                                         & 0.8724                          & 0.9256                                  \\ \bottomrule
    \end{tabular}
    \label{tab:giants-tank-test-results}
\end{table}

\begin{figure}[]
\centering
\subcaptionbox{}{\includegraphics[width=0.3\textwidth]{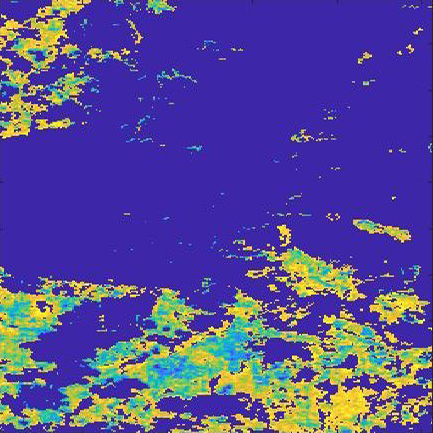}}%
\hfill 
\subcaptionbox{}{\includegraphics[width=0.3\textwidth]{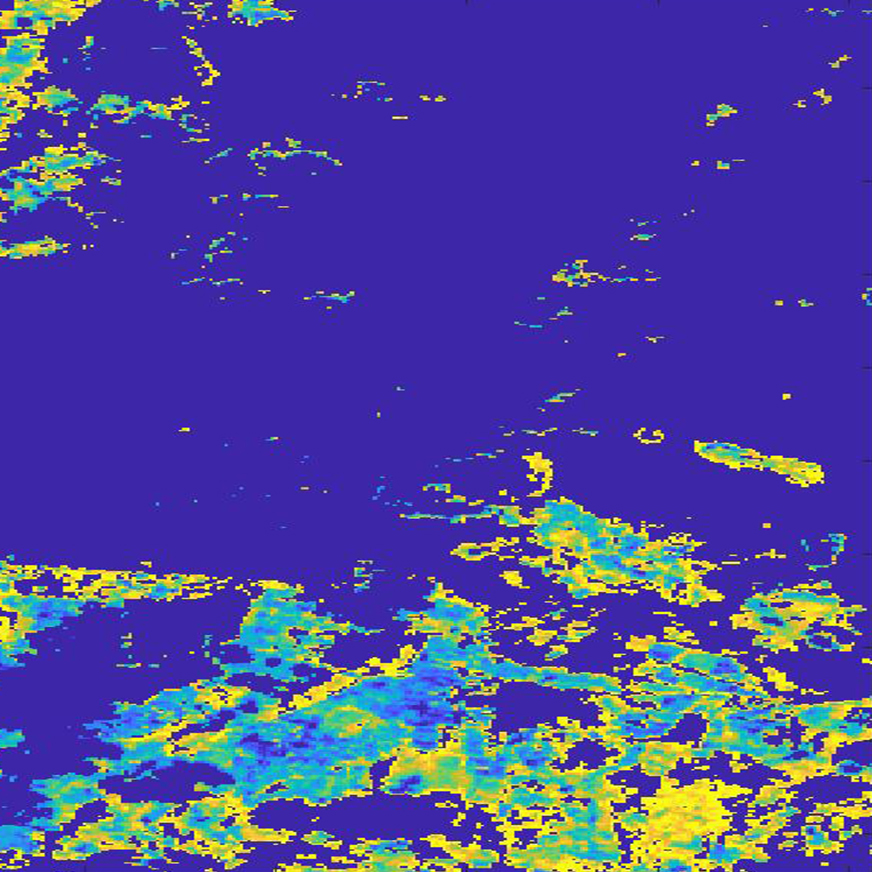}}%
\hfill 
\subcaptionbox{}{\includegraphics[width=0.3\textwidth]{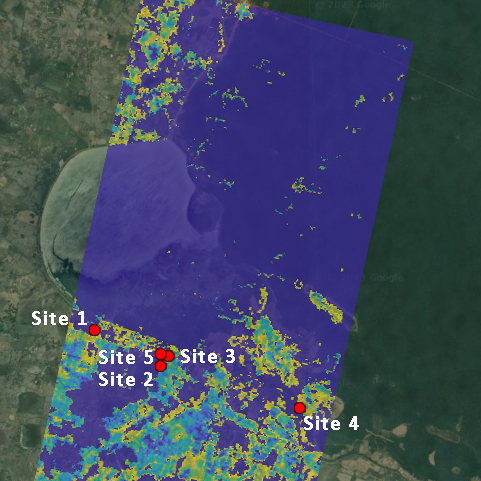}}%
\caption{(a) Relative availability Map (b) Abundance map generated after NNLS (Alpha values generated after NNLS) (c) Generated mineral map overlapped with sites at Giant's Tank}
\label{figure:Giants-tank-relative-availability}
\end{figure}

The abundance map generated by the algorithm (Fig. \ref{figure:Giants-tank-relative-availability}) indicates the widespread presence of Montmorillonite throughout the region. This observation can be attributed to the extensive coverage of paddy fields supplied by the Giant’s Tank. To validate the sites recommended by the abundance values ($\alpha$), four specific sites were selected for mineral availability testing using XRD. The results of the XRD tests, along with the abundance predictions from the algorithm, are presented in table \ref{tab:giants-tank-test-results}. Site 3, with the highest Montmorillonite percentage, aligns with the algorithm's prediction, accurately identifying it as having the highest abundance. 

The algorithm has shown high abundant values for sites 3 and 5 while showing intermediate abundance values  for the other sites. The calculated correlation coefficient between the XRD test results and the relative availability was 0.6504. This value could be attributed to the fact that the pixels may have had a large cloud coverage which resulted in the algorithm to perform imprecisely for certain sites. However, similar to the other cases, the generated abundance values shows a good correspondence with the XRD test results.

\section{Conclusion}
\label{section: conclusion}

The task of detecting and mapping a specific mineral amidst impurities presents a formidable challenge within the domain of Digital Mineralogical Mapping (DLM). The difficulty stems from the need to distinguish the target mineral from surrounding impurities accurately. In response to this problem, this study proposed an effective solution.

In summary, this research has yielded a novel algorithm capable of delivering accurate results, regardless of the particular mineral of interest and the varying mineral concentrations in the presence of impurities. The algorithm's versatility and effectiveness are demonstrated by its impressive results across the four selected study locations, extensively discussed in the results and discussion section.

Within the algorithm's framework, a novel autonomous approach of separating mineral and impurity representative subclasses was introduced which could be used as an alternative in place of traditional heuristic approaches. Followed by this, an intermediate step involved extraction of mineral representative signature from the soil manifold of the and deriving the eigen-direction that optimally separates mineral pixels from impurities based on the sub classes derived. This approach facilitated the calculation of a critical parameter known as "relative availability," which is instrumental in generating precise mineral maps. Additionally, the incorporation of non-negative least squares estimation played a vital role in determining mineral availability. These results were rigorously compared with on-site data obtained through XRD and Magnetic Separation tests. 

Furthermore, this study addressed a common issue in Digital Mineralogical Mapping, which is the reduced accuracy associated with relying solely on laboratory reference signatures. It addressed this challenge by identifying mineral representative signatures specific to the regions under study. This allows to capture the relative availability of mineral with respect to the mineral representative signature inherent to the region. 

The authors are optimistic about the global applicability of this algorithm, as the study encompassed internationally significant minerals. This algorithm has the potential to streamline mineral surveys, making the identification of potential mineral deposits more efficient. Looking ahead, the authors aim to enhance this methodology by integrating deep learning models, paving the way for further advancements in the field of Digital Lithological Mapping.




\section*{Acknowledgment}
The authors acknowledge the support from the USGS for providing hyperspectral images from the EO-1 satellite's Hyperion sensor, and documents for standard preprocessing of the images. We are grateful to the Department of Geology, University of Peradeniya for arranging and providing equipment for soil sample preparation.  Also, Prof. H.M.V.R. Herath likes to acknowledge the supported in part at the Technion by a fellowship from the Lady Davis Foundation. The authors acknowledge the support received from J.M.V.D.B. Jayasundara, H.M.K.D. Wickramathilaka and N. Senarath during the results generation, and Mr. Nevil Attanayake and Mr. Chanaka Perera during the fieldwork, and Mr. Kamal Jayasinghe of the Postgraduate Institute of Science, University of Peradeniya for conducting the laboratory tests.

\bibliographystyle{tfcad}
\bibliography{main}

\end{document}